\documentclass[12pt,preprint]{aastex}

\begin{document}
\title{Highly Ionized High-Velocity Clouds toward PKS 2155$-$304 
   and Markarian 509}
\author{Joseph A. Collins, J. Michael Shull\altaffilmark{1}}
\affil{University of Colorado, CASA, Department of Astrophysical \& 
     Planetary Sciences, Campus Box 389, Boulder, CO 80309}
\altaffiltext{1}{Also at JILA, University of Colorado and National 
     Institute of Standards and Technology.}
\and
\author{Mark L. Giroux}
\affil{East Tennessee State University, Department of Physics \& 
     Astronomy, Box 70652, Johnson City, TN 37614}

\begin{abstract}

To gain insight into four highly ionized high-velocity clouds (HVCs) 
discovered by Sembach et al.\ (1999), we have analyzed data from the
{\it Hubble Space Telescope} ({\it HST}) and {\it Far Ultraviolet
Spectroscopic Explorer} ({\it FUSE}) for the PKS 2155$-$304 and Mrk 509
sight lines.  We measure strong absorption in \ion{O}{6} and
column densities of multiple ionization stages of silicon (Si~II/III/IV)
and carbon (C~II/III/IV).  We interpret this ionization pattern as a 
multiphase medium that contains both collisionally ionized and photoionized 
gas.  Toward PKS~2155$-$304, for HVCs at $-140$ and $-270$ km~s$^{-1}$, 
respectively, we measure log N(\ion{O}{6}) $= 13.80\pm 0.03$ and 
log N(\ion{O}{6}) $= 13.56 \pm 0.06$; from Lyman series absorption, we 
find log N(\ion{H}{1}) $= 16.37^{+0.22}_{-0.14}$ and 
$15.23^{+0.38}_{-0.22}$.  
The presence of high-velocity
\ion{O}{6} spread over a broad (100 km~s$^{-1}$) profile, together with 
large amounts of low-ionization species, is difficult to reconcile with 
the low densities, $n_e \approx 5 \times 10^{-6}$ cm$^{-3}$, in the 
collisional/photoionization models of Nicastro et al.\ (2002), although 
the HVCs show a similar relation in $N$(\ion{Si}{4})/$N$(\ion{C}{4}) 
versus $N$(\ion{C}{2})/$N$(\ion{C}{4}) as high-$z$ intergalactic clouds. 
Our results suggest that the high-velocity \ion{O}{6} in these absorbers
do not necessarily trace the WHIM, but instead may trace HVCs with low
total hydrogen column density. 
We propose that the broad high-velocity 
\ion{O}{6} absorption arises from shock ionization, at bowshock 
interfaces produced from infalling clumps of gas with velocity shear.   
The similar ratios of high ions for HVC Complex C and these highly ionized 
HVCs suggest a common production mechanism in the Galactic halo.

\end{abstract}
 

\section{INTRODUCTION}

The high-velocity clouds (HVCs; Wakker \& van Woerden 1997) seen in 
\ion{H}{1}\ 21 cm emission are characterized by significant deviation
from a simple model of differential Galactic rotation.  Although they are 
thought to reside in the Galactic halo, their origin within the context
of galaxy assembly is not well understood.  The fact that a large fraction 
of HVCs exhibit negative velocity centroids in the 
local standard of rest (LSR), 
and the need to explain the metallicity distribution of long-lived 
stars in the solar neighborhood (Flynn \& Morell 1997), have motivated the 
suggestion that these HVCs represent the accretion of low-metallicity gas 
onto the Galaxy (Wakker et al.\ 1999; Collins, Shull, \& Giroux 2003, 
hereafter CSG).  Negative velocity HVCs may be condensed remnants of a 
``Galactic fountain'' (Shapiro \& Field 1976; Bregman 1980) now returning 
to the disk.  

An alternate, but controversial hypothesis places some of the HVCs at 
locations in the Local Group (Blitz et al.\ 1999), perhaps 500 kpc distant, 
and identifies the compact HVCs (CHVCs) in particular as a population of 
dwarf-type galaxies at large distance (Braun \& Burton 1999).  However, 
evidence
supporting this hypothesis has not been forthcoming, since both a stellar
component to the CHVCs (Simon \& Blitz 2002; Hopp, Schulte-Ladbeck, \&
Kerp 2003) and similar \ion{H}{1}\ clouds in nearby groups (Zwaan 2001)
have yet to be detected.  Recent H$\alpha$ detections of HVCs and CHVCs
indicate that many of these objects are within a $\sim40$ kpc radius of
the Galaxy (Putman et al. 2003).  In addition, models suggest that many 
CHVCs at Local Group distances would have unreasonably large total masses,
summing the neutral, ionized, and dark matter (Maloney \& Putman 2003). 
This discussion of HVC location within the Galaxy halo or Local Group
has expanded to include a population of \ion{O}{6} HVCs (Sembach et al.\ 
2003; Nicastro et al.\ 2003).  
In a specific instance, abundance studies of HVC Complex C with the 
{\it Far Ultraviolet Spectroscopic Explorer} ({\it FUSE}) and 
{\it Hubble Space Telescope} ({\it HST}) suggest that this gas is an 
infalling low-metallicity cloud mixing with gas of Galactic origin 
(CSG; Tripp et al.\ 2003). 

An additional consideration is the HVC hot gas content.
Cosmological simulations predict that the collapse of gas during galaxy 
formation should produce significant amounts of shock-heated gas
(Cen \& Ostriker 1999) in the low-redshift intergalactic medium (IGM),
the so-called warm-hot IGM or WHIM.  Temperatures of this  
shocked gas would be $10^{5}-10^{7}$~K, resulting in significant ionization 
fractions of \ion{O}{6}, \ion{O}{7}, and, if the temperature is 
sufficiently high, \ion{O}{8}.  A goal of recent {\it FUSE} surveys of 
high-velocity \ion{O}{6} has been to investigate the interaction of HVCs 
with the WHIM and hot gas in the Galactic halo (Sembach et al. 2003; 
Indebetouw \& Shull 2004).  If HVCs are remnants
of Local Group formation, then the associated \ion{O}{6} 
may trace collisionally ionized gas at the cloud/IGM interface.  

In this paper, we use new data from the E140M echelle on the 
{\it Space Telescope Imaging Spectograph} (STIS) to assess the physical 
characteristics of a class of ``highly ionized HVCs".  These clouds
were discovered by Sembach et al.\ (1995) and further defined by 
Sembach et al.\ (1999, hereafter S99).  Using data from the {\it Goddard 
High Resolution Spectrograph} (GHRS), S99 studied four of these HVCs along 
the lines of sight toward the AGN PKS 2155$-$304 and Mrk 509. These
absorbers exhibited strong \ion{C}{4}\ absorption accompanied by little or 
no absorption in low ions.  High-velocity \ion{H}{1}\ 
emission was detected in an area encompassing $\sim2\arcdeg$ near the sight
lines, but no \ion{H}{1}\ emission could be detected in the sight lines 
themselves down to sensitivity levels of 
$N$(\ion{H}{1})$\gtrsim10^{18}$ cm$^{-2}$.  
In addition, \ion{H}{1} emission associated with the Galactic Center
Negative (GCN) clouds can be detected at similar velocities to the HVCs 
within $\sim5\arcdeg$ of the sight lines, indicating a possible association
between the HVCs and the GCN clouds. 
The column density constraints from the GHRS data were successfully modeled 
as arising from an extragalactic QSO-type radiation field.  
The resulting densities,
pressures, and sizes of the clouds suggested to S99 that these HVCs were
most likely located in the Local Group or distant Galactic Halo, and
that the highly ionized gas traces the outer, low-density regions of the HVCs.  

However, \ion{O}{6} has recently been detected in these HVCs (Sembach et al.\
2003) at column densities well above any reasonable prediction from photoionization
models, indicating a contribution from other ionization sources.  
Nicastro et al.\ (2002) detected zero-redshift \ion{O}{7} X-ray absorption 
in the PKS 2155$-$304 sight line, and identified it as a filament of
WHIM.  Because the resolution of the X-ray data is poor ($\sim600$ 
km~s$^{-1}$ at 20 \AA) relative to the UV data, it is not possible to 
distinguish Galactic absorption from the HVC absorption in those data.  

Nevertheless, Nicastro et al. (2002) model the X-ray absorption lines
{\it and} the high-velocity \ion{O}{6} absorption as arising in the
same low-density ($n_e \approx 5 \times 10^{-6}$ cm$^{-3}$)
collisionally ionized WHIM component.  They suggest that the clouds
are primarily collisionally ionized, with a contribution from
photoionization by the extragalactic UV/X-ray background.  This
interpretation requires a large photoionization parameter, $U =
n_{\gamma}/n_H$, and a low gas density, $n_H \approx 5\times10^{-6}$
cm$^{-3}$. The large implied cloud sizes suggest that the PKS
2155$-$304 HVCs are at Local Group distances.  This hypothesis has
important astronomical consequences. The FUSE survey of HVCs (Sembach
et al.\ 2003) found high-velocity \ion{O}{6} over a covering fraction
$f_{\rm sky} = 0.60-0.85$, down to columns log~N(\ion{O}{6}) = 13.95.
For an assumed metallicity, (O/H)$=5.45\times10^{-5}$ (Holweger 2001), 
that is 10\% of solar,
and $T=10^{5.45}$ K gas
temperature (ionization fraction, $f_{\rm OVI} = 0.2$), this
corresponds to an ionized hydrogen column density N(H$^+) = 8 \times
10^{18}$ cm$^{-2}$.  This implies a total gas mass $M_{\rm hot}
\approx (10^{10}~M_{\odot}) f_{\rm sky} d_{100}^2$ at mean distance
(100~kpc)$d_{100}$. Placing the \ion{O}{6} HVCs at $d \sim 1$~Mpc
would then result in masses exceeding that of the entire Milky Way
galaxy.

Heckman et al.\ (2002) also cast doubt on the ``Local Group" interpretation 
by pointing out that the entire population of \ion{O}{6}\ absorbers, including 
those of known IGM clouds, can be modeled as ``cooling layers" of gas,
with initial temperature $T \approx 10^{6}$ K.  In such a non-photoionization 
scenario, the large cloud sizes and low densities are unnecessary, and the 
HVCs could instead trace gas in the Galactic halo.  

With the recent acquisition of {\it HST}/STIS echelle spectra for both
PKS~2155$-$304 (Shull, et al.\ 2003) and Mrk~509
(Kraemer et al.\ 2003) comes the opportunity to reexamine the
highly-ionized HVCs toward these targets.  The STIS has superior
sensitivity, resolution, and spectral coverage to GHRS, allowing a
more thorough investigation of a wider variety of absorption lines.
In addition to providing important information on \ion{O}{6},
\ion{N}{2}, and \ion{C}{3}, {\it FUSE} spectra include coverage of
\ion{H}{1} absorption lines from Ly$\beta$ down to the hydrogen Lyman
limit (Moos et al.\ 2000).  If the data quality is high, the higher
Lyman lines can be used to determine $N$(\ion{H}{1}) for the HVCs,
which is critical for restricting parameter space in ionization
models.  For PKS 2155$-$304, we are able to measure $\log
N$(\ion{H}{1}) of the HVCs to $\pm0.2-0.3$ dex.

Using measured column densities of multiple ion species, we investigate the 
physical conditions and characteristics of four of these highly-ionized HVCs
toward two AGN.  In light of 
the ``Local Group hypothesis" for HVC distances, we attempt to determine 
whether these clouds are extragalactic objects or associated with the Galactic 
halo. Because of the substantial O~VI column densities detected, we 
suggest that O~VI and other high ions arise in collisionally ionized gas.   
In addition, the presence of highly-ionized species allows us to
investigate the interaction of HVCs with the IGM or Galactic halo.  
Our method of processing of the raw data is discussed in \S\ 2.  The column
density measurements are presented in \S\ 3.  In \S\ 4, we discuss the results, 
present models for the ionization of these clouds, and compare 
the observed column density ratios to measurements of multiple ions of
carbon (C~II/III/IV) and silicon (Si~II/III/IV) in the IGM 
at high and low redshift.

\section{FUSE AND HST-STIS DATA}

Data for these sight lines consist of {\it HST}/STIS  echelle and {\it FUSE} 
spectra obtained from the Multimission Archive (MAST) at the Space 
Telescope Science Institute.  The {\it HST} data for PKS~2155-304 were 
obtained in a ten-orbit STIS exposure as part of GO Program 8125 in
Cycle 8 (Shull et al.\ 2003). The {\it FUSE} data were 
obtained in observations by the {\it FUSE} Science Team.   
A summary of the observations is shown in Table 1.

\placetable{t1}

Calibrated {\it FUSE} spectra were extracted by passing raw data 
through the CALFUSE Version 2.2 reduction pipeline.  In order to improve
the signal-to-noise ratio (S/N), both ``day'' and ``night'' 
photons are included in the final calibrated {\it FUSE} spectra.
The inclusion of day photons leads to strong airglow contamination of
certain neutral interstellar lines near the LSR.  
Such airglow contamination is irrelevant for the purpose
of investigating high-velocity absorption in this study.  Individual 
exposures were then co-added and weighted by their exposure times, to yield a
final {\it FUSE} spectrum.  To further improve S/N, the data were rebinned over
5 pixels.  {\it FUSE} spectra have a resolution over the FUSE bandpass (912-1187 \AA) of about 20 km s$^{-1}$ 
($\sim10$ pixels), and as a result the data are oversampled at that resolution.  
To set the absolute wavelength scale, the centroid of Galactic \ion{H}{1}\
21-cm emission was compared and aligned
to various Galactic absorption lines in the {\it FUSE}
bandpass, such as those of \ion{Si}{2}\ ($\lambda$1020.70), 
\ion{O}{1}\ ($\lambda$1039.23), \ion{Ar}{1}\ ($\lambda\lambda$1048.22, 
1066.66), \ion{Fe}{2}\ ($\lambda\lambda$1125.45, 1144.94), \ion{N}{1}\
($\lambda$1134.17), and various H$_{2}$ Lyman bands.  

The {\it HST}/STIS observations were taken in the E140M echelle mode, providing
a wavelength coverage of $1150-1700$ \AA\ at 7 km s$^{-1}$ resolution.  A 
final STIS spectrum was obtained by co-adding individual exposures,
weighted by their exposure times and rebinning the data over 3 pixels.   An 
absolute wavelength scale was obtained as for the {\it FUSE} spectrum, matching
the Galactic \ion{H}{1}\ emission peak to absorption features of \ion{N}{1}\
($\lambda$1199.55), \ion{S}{2}\ ($\lambda\lambda\lambda$1250.58, 1253.80, 
1259.52), \ion{O}{1}\ ($\lambda$1302.17), and \ion{Si}{2}\ 
($\lambda\lambda$1304.37, 1526.71). We estimate the absolute wavelength scales 
of both the STIS and {\it FUSE} data to be accurate to within $\sim10$ km s$^{-1}$.
   
\section{COLUMN DENSITY MEASUREMENTS}

In both the PKS 2155$-$304 and Mrk 509 sight lines, high velocity metal-line
absorption is clearly present (see Figures 1 and 6).  
In order to ascertain characteristics and 
physical conditions in these clouds, it is important to determine column
densities of the various ion species.  In the few 
cases where multiple lines of one 
species can be measured, we use a curve-of-growth (CoG) 
fitting procedure to measure 
column densities.  More commonly, we measure columns with the apparent 
optical depth (AOD) method.  Though a CoG fit is the more robust method of 
determining column densities, the AOD method is valid for 
unsaturated lines (Savage \& Sembach
1991).  Line saturation can be difficult to detect, particularly if such 
saturation is unresolved.  However, we can test for the presence of unresolved 
saturation in the case of doublets (e.g., the \ion{C}{4} and \ion{Si}{4}
doublets) where the implied optical depths of the weak and strong lines
can be compared.  
In cases of saturated doublets we use a CoG fit to determine the column 
density.
The optical depth of 
absorption at velocity $v$ is given by,
\begin{equation}
\tau_{v} = - {\rm ln}[I_{n}(v)]   \;  , 
\end{equation}
where $I_{n}(v)$ is the observed normalized intensity at velocity $v$.  
Individual line profiles
were normalized by fitting low-order polynomials to the continuum immediately
surrounding the line in question.  Typically, we fit the continuum $\pm3$-5 
\AA\ about the rest wavelength of the line,  although in a number 
of cases spurious absorption near the line required the use of a much 
larger region for continuum measurement.  The column density is then
given by integrating $\tau_{v}$ over the velocity range of the absorption 
feature,
\begin{equation}
N_{AOD}({\rm cm^{-2}}) = \left(\frac{3.767\times10^{14}}{f\lambda}\right) 
      \sum_{v}\tau_{v }{\Delta}v   \;  , 
\end{equation}
where ${\Delta}v$ is the bin size in km s$^{-1}$ and values of the 
oscillator strength, $f$, and the rest wavelength, $\lambda$ (in \AA), are 
from D. C. Morton (2004, in preparation). 
Measurements for the HVCs in the PKS 2155$-$304 and Mrk 509 sight lines 
are discussed individually in the following sections.

\subsection{PKS 2155$-$304}

Absorption-line profiles of metal ions in the {\it FUSE} and STIS 
bandpass are shown in Figure 1.  As in the GHRS data presented by S99, 
the profiles show two distinct high-velocity components. The
stronger component occupies the velocity range of $-200<V_{LSR}<-95$ km 
s$^{-1}$ (centered at $-140$ km s$^{-1}$) while the other is 
located in the range $-310<V_{LSR}<-200$ km s$^{-1}$ (centered at $-270$
km s$^{-1}$).  Each of the components show highly ionized absorption in 
\ion{C}{4}, \ion{Si}{4}, and \ion{O}{6}, while only the $-140$ km s$^{-1}$
component shows singly-ionized absorption in \ion{C}{2}\ and \ion{Si}{2}.
The extent of the singly ionized and highly ionized absorption lines are
noticeably different for the $-140$ km s$^{-1}$ component.  Figure 2
shows normalized absorption profiles of carbon and silicon species.  Lines of 
\ion{C}{2}\ and \ion{Si}{2}\ are confined to the range $-165<V_{LSR}<-95$ km 
s$^{-1}$, 
while absorption from the more highly-ionized species extends to higher 
velocities.  This difference hints at a possible 
multi-component structure to the $-140$ km s$^{-1}$ component, with the 
highly-ionized species tracing a more extended halo 
than the less ionized core.  The profile of
\ion{C}{3}\ absorption may be a reflection of saturation instead of a truly
flat column density distribution. 

\placetable{t2}

Table 2 shows measured equivalent widths and column densities for each of the
components in the PKS 2155$-$304 sight line.  For the $-140$ km s$^{-1}$ 
component, the measurement of multiple lines of \ion{Si}{2}\ and \ion{C}{2}
allows us to use a CoG to measure column densities, or their 
upper limits for the neutral and singly ionized metal species. 
The three \ion{Si}{2}\ and two \ion{C}{2}\ lines are best fitted by a CoG with
doppler parameter $b=10.3$ km s$^{-1}$, as shown in Figure 3.  The profiles
of \ion{C}{3}\ and \ion{Si}{3}\ for the $-140$ km s$^{-1}$ component show 
evidence of possible saturation and, as a result, their measured column
densities place lower limits on the actual columns.   
Within the errors, the column densities presented in Table 2
are consistent with the columns derived from GHRS data presented by
S99.  The quality of the GHRS data led S99 to conclude, however, that
these clouds show little or no detectable low ion absorption.  Our results 
indicate measurable columns of low-ionization species for the $-140$ km
s$^{-1}$ component (in \ion{H}{1}, \ion{C}{2}, and \ion{Si}{2}), as well as 
for the $-270$ km s$^{-1}$ component (in \ion{H}{1}, but only upper limits
on \ion{C}{2} or \ion{Si}{2}).

\placetable{t3}

The {\it FUSE} SiC channel data quality is sufficient to obtain 
equivalent widths for many \ion{H}{1}\ Lyman-series absorption lines.
Profiles of \ion{H}{1}\ absorption lines in this sight line are shown in
Figure 4.  The \ion{H}{1} column of the $-270$ km s$^{-1}$ component is 
sufficiently low that unsaturated absorption can be detected in Ly$\gamma$, 
with the component 
eventually becoming undetectable at Ly$\theta$.  Owing to blending with
saturated Galactic \ion{H}{1}\ absorption, only lines from Ly$\eta$ through
Ly$\lambda$ are useful for measuring the \ion{H}{1}\ column of the 
$-140$ km s$^{-1}$ component.  Equivalent width measurements of the \ion{H}{1}\
Lyman lines for the HVC components are shown in Table 3, for 
four and five \ion{H}{1}\ lines for the $-270$ 
and $-140$ km s$^{-1}$ components, respectively.  
We employed a CoG analysis (Fig.\ 5) to determine the 
corresponding \ion{H}{1}\ column densities of these components, 
log$N$(\ion{H}{1})$=16.37^{+0.22}_{-0.14}$ for the $-140$ km s$^{-1}$ 
component and log$N$(\ion{H}{1})$=15.23^{+0.38}_{-0.22}$ for the $-270$ km 
s$^{-1}$ component.  This is a significant improvement over the work of S99, 
where only an upper limit could be determined based on 
single-dish \ion{H}{1}\ emission data.  These low \ion{H}{1} HVC column 
densities show why no 21-cm emission was detected.

\subsection{Mrk 509} 

Profiles of metal ion absorption lines observed 
in the {\it FUSE} and STIS 
data are shown in Figure 6.  The data quality is not as good as that 
for PKS 2155$-$304, but high-velocity absorption can be detected.
As with PKS 2155$-$304, the profiles show 
two distinct high-velocity components. 
In the Mrk 509 sight line, however, 
the stronger component appears at higher velocity, 
$-355<V_{LSR}<-260$ km s$^{-1}$ (centered at $-300$ km s$^{-1}$),
than the weak component, which can 
only be detected at the 3$\sigma$ level in a few absorption lines.
The weak component is most obvious in \ion{C}{4} $\lambda1548.20$ and 
\ion{Si}{3} $\lambda1206.50$.  These lines 
are used to establish an 
integration range of $-260<V_{LSR}<-200$ km s$^{-1}$
(centered at $-240$ km s$^{-1}$) for the weak component.  
The \ion{O}{6} $\lambda1031.93$ absorption from the $-240$ km s$^{-1}$
component is contaminated by weak absorption from the L(6-0) P(3) Galactic
H$_{2}$ line.  The nearby L(7-0) P(3) line has a similar oscillator
strength to the L(6-0) P(3) line.  To remove this contamination, 
we inferred the optical depth of the contaminating line by scaling the 
integrated $\tau_{v}$ 
of the Galactic H$_{2}$ L(7-0) P(3) line by its relative value of $f\lambda$.
The line width of the Galactic H$_{2}$ L(7-0) P(3) is then used to subtract
off the appropriate integrated $\tau_{v}$ for the 
contaminating absorption.  The resulting \ion{O}{6}\ profile
looks reasonable, although the removal of the H$_{2}$ line has effectively 
smoothed over any possible small-scale structure. 

The {\it FUSE} SiC channel data are so poor 
that we could not obtain measurements
of lines below 1000 \AA\ or between 1080$-$1100 \AA.  Thus, our results do not
include column density measurements based on \ion{N}{2} $\lambda1083.994$, 
\ion{C}{3} $\lambda977.020$, or the \ion{H}{1} Lyman lines.  The lack of an
\ion{H}{1} measurement is perhaps most critical, and we were unable to improve 
on the upper limit from single-dish \ion{H}{1} emission data 
of log$N$(\ion{H}{1})$<17.69$ established by Sembach et al.\ (1995).    

\placetable{t4}

Measured equivalent widths and column densities for each of the high-velocity
components in the Mrk 509 sight line are shown in Table 4.  The column density
pattern is similar to the HVCs in the PKS 2155$-$304 sight line.  
The core of the \ion{Si}{3}\ $\lambda1206.500$ profile for the 
$-300$ km s$^{-1}$ component is likely saturated, 
and the measured column density should be considered a lower limit.
Although saturation is not 
apparent in the \ion{C}{4}\ and \ion{Si}{4}\ profiles,
it is possible that unresolved saturation can be an issue, since the weak
lines give $\sim0.1$ dex larger column densities for the
$-300$ km s$^{-1}$ component.
In order to test for unresolved saturation, we compared the profiles 
of the weak and strong lines in the \ion{C}{4}\ and \ion{Si}{4}\ doublets. 
If saturation were an issue, the AOD method should give a larger and more
peaked column density for the weak line than for the optically thicker 
strong line.  Figure 7 shows the normalized column density profiles of the 
\ion{C}{4} and \ion{Si}{4} doublets for the $-300$ km s$^{-1}$ component.
Clearly, unresolved saturation may be present, since
the profiles from the weaker lines
of the doublets are more peaked between $-300$ and $-280$
km s$^{-1}$.  Due to the presence of saturation, a CoG fit to the doublets
is used to measure 
the column densities of \ion{C}{4} and \ion{Si}{4}.
Since the feature at $-350$ km s$^{-1}$ in the \ion{Si}{4}\ 
$\lambda1402.770$ profile is not observed in the \ion{Si}{4}\
$\lambda1393.755$ profile, we suspect that this feature is a result of 
spurious absorption in the integration range. Therefore, absorption from
this feature is not included in the $W_{\lambda}$ measurement of 
\ion{Si}{4}\ $\lambda1402.770$.  

\section{DISCUSSION}
\subsection{Column Density Ratios}

An obvious characteristic of these HVCs is their strong absorption in
highly-ionized species (C~IV, Si~IV).  Furthermore, except for the
$-140$ km~s$^{-1}$ component (PKS~2155$-$304), the column densities of
singly-ionized and neutral species are quite low or below the
detection limit.  There are few HVCs that have been similarly studied
with both {\it FUSE} and {\it HST}/STIS spectra.  One well-studied
case is the sight line toward PG 1259+593, which probes the extended
HVC Complex C (Richter et al. 2001; CSG).  We have gone back to the
data for PG 1259+593 and measured the column densities of the common
high ions\footnotemark.  \footnotetext{The column densities of
highly-ionized species associated with Complex C in the PG 1259+593
sight line are: log$N$(\ion{C}{4})$=13.23^{+0.06}_{-0.08}$;
log$N$(\ion{N}{5})$<13.16$;
log$N$(\ion{O}{6})$=13.54^{+0.05}_{-0.05}$;
log$N$(\ion{Si}{4})$=12.64^{+0.10}_{-0.13}$.  The measurements were
made with the AOD method using the techniques described in CSG.}
Although the \ion{H}{1} column density of Complex C derived from
Effelsberg single-dish 21-cm data (9$\arcmin$ beam size) along that
sight line is several orders of magnitude larger than for the highly
ionized HVCs,
log$N$(\ion{H}{1})$=19.92^{+0.01}_{-0.01}$ (Wakker et al. 2001), the
measured column densities of \ion{C}{4}, \ion{Si}{4}, and \ion{O}{6}\
are within an order of magnitude of those for the PKS 2155$-$304 and Mrk 509
HVCs.

\placetable{t5}

Table 5 perhaps best illustrates the similarities and differences between
these highly-ionized HVCs and Complex C, listing various 
logarithmic column density ratios.  In low-to-high ion ratios, 
the highly-ionized HVCs have values
smaller than Complex C by $\sim10^{3}$ to $10^{4}$ in 
$N$(\ion{H}{1})/$N$(\ion{O}{6}) and $\sim10^{2}$ in 
$N$(\ion{Si}{2})/$N$(\ion{Si}{4}).
However, the highly-ionized HVCs and Complex C show similar ratios involving 
only the high ions.
An exception is the ratio $N$(\ion{C}{4})/$N$(\ion{Si}{4}) where the value
for Complex C is lower than for the highly-ionized HVCs.
The value for Complex C, 
log[$N$(\ion{C}{4})/$N$(\ion{Si}{4})]$=0.59^{+0.12}_{-0.15}$, 
is similar to that observed for gas within the disk or low halo, 
log[$N$(\ion{C}{4})/$N$(\ion{Si}{4})]$=0.58^{+0.18}_{-0.30}$
(Sembach, Savage, \& Tripp 1997), whereas the values 
for the highly-ionized HVCs are slightly larger.
In addition, the weaker of the highly-ionized HVC 
components (the $-270$ km s$^{-1}$ component toward PKS 2155$-$304 and the 
$-240$ km s$^{-1}$ component toward Mrk 509) show larger values of
$N$(\ion{C}{4})/$N$(\ion{Si}{4}) than for the stronger components.
Interestingly, the ratio $N$(\ion{O}{6})/$N$(\ion{C}{4}) shows
little difference between the highly-ionized HVCs and Complex C, 
as well as between the individual HVC components themselves.  
Fox et al. (2004) find that the high-ion 
ratios in Complex C are consistent with ionization at conductive or turbulent
interfaces between the cloud core and a surrounding hot medium.
The similar high-ion ratios for the highly ionized HVCs and Complex C may 
indicate that these objects share a common production mechanism for the 
\ion{C}{4} and \ion{O}{6}.   

\subsection{Comparisons with the IGM} 

As demonstrated in the previous section, the column density ratios
involving the low-ions are unlike those of gas normally associated
with the Galaxy.  In this section, we compare the observations to
those of the IGM.  Characteristics of IGM gas are diverse, and we can
compare these observations to absorbers at both high- and
low-redshift.  The values of $N$(\ion{Si}{4})/$N$(\ion{C}{4}) in these
HVCs are similar to values typical for Ly$\alpha$ forest clouds at $z
\approx 3$, for example those studied by Songaila \& Cowie (1996;
hereafter SC).  Figure 8 shows a diagnostic diagram of
$N$(\ion{Si}{4})/$N$(\ion{C}{4}) versus
$N$(\ion{C}{2})/$N$(\ion{C}{4}) for the SC data.  The large circles
denote data for the four HVC components observed in the PKS 2155$-$304
and Mrk 509 sight lines.  Clearly, the HVCs show values of
$N$(\ion{Si}{4})/$N$(\ion{C}{4}) similar to those of the SC sample.
The same can be said for the $N$(\ion{C}{2})/$N$(\ion{C}{4}) ratio for
three of the HVC components.  The $-140$ km s$^{-1}$ component towards
PKS 2155$-$304, however, has a value of
$N$(\ion{C}{2})/$N$(\ion{C}{4}) more than twice as large as any of the
Ly$\alpha$ forest clouds in the SC sample.  Though the value of
$N$(\ion{Si}{4})/$N$(\ion{C}{4}) for this HVC is similar to that of
the $z\sim3$ Ly$\alpha$ forest, it contains higher fractions of
low-ionization species (\ion{C}{2}) than typically associated with the
IGM.

At low-redshift, Tripp et al.\ (2002) observed metal ions in two Ly$\alpha$
absorbers in the 3C~273 sight line at redshifts in the vicinity 
of the Virgo cluster.  One of the absorbers shows metal absorption in only 
\ion{O}{6}\ and is identified by the authors as possibly tracing the WHIM.  
The other absorber shows absorption in \ion{Si}{2}, \ion{Si}{3}, and \ion{C}{2}, 
although no absorption is seen in higher ionization species.  The authors identify
this cloud as a possible Virgo analog to HVCs near the Milky Way.  This 
absorber has a lower limit, log [N(\ion{C}{2})/N(\ion{C}{4})] $> 0.24$, 
indicating that it may have similar properties to the $-140$ km~s$^{-1}$ 
component towards PKS 2155$-$304, but at lower \ion{H}{1} column density. 

Nicastro et al.\ (2002, 2003) interpret the high-velocity \ion{O}{6}
(FUSE) detection towards PKS 2155$-$304 and many other sight lines as
tracing the WHIM in the Local Group.  Their hypothesis is based on the
detection of O~VII/O~VIII absorption at $z=0$ and the narrowing of the
O~VI velocity distribution when converted to a Local Group standard of
rest.  Possible detections of the WHIM at higher redshifts have been
reported in \ion{O}{7} and \ion{O}{8} (Fang et al. 2002; Mathur,
Weinberg, \& Chen 2003), although many of those results have not been
confirmed (Rasmussen, Kahn, \& Paerels 2003).  The high-velocity
\ion{O}{6}\ in the Nicastro et al.\ picture would be at high
temperature ($T\gtrsim10^{5}$ K), and significant amounts of
\ion{O}{7} and \ion{O}{8} could be present, which are not only
difficult to detect but difficult to resolve in X-ray
data. Nevertheless, the presence of large amounts of low ionization
species in the PKS 2155$-$304 HVCs is incompatible with the low
densities in the collisional/photoionized models of Nicastro et al.\ (2002).  
This
implies, at least for the HVCs towards PKS 2155$-$304 and Mrk 509,
that the high-velocity \ion{O}{6} do not necessarily trace the WHIM, but
instead may trace HVCs with low total hydrogen column density.
It is possible that some of the \ion{C}{4}, \ion{Si}{4}, and lower
ions could arise from thermal instability in cooling WHIM
gas. However, this scenario may be inconsistent with the broad
\ion{O}{6} velocity profiles.

\subsection{Modeling the Ionization Pattern in HVCs}

As indicated above, the HVCs toward PKS 2155$-$304 and Mrk 509 show column 
density ratios unlike gas associated with the Galactic halo.  Although the 
\ion{H}{1}\ column densities are considerably higher, Galactic halo gas 
typically has much stronger absorption from low ions than high ions
(Sembach \& Savage 1992).  These HVCs are also distinguished (S99) from 
Galactic or low halo gas by their high values of 
N(\ion{C}{4})/N(\ion{Si}{4}) $> 5$.  In order to examine the ionization 
characteristics of these HVCs and assess their possible origin, we have
performed a series of model calculations which are then constrained 
by the observed column densities.  The HVCs toward PKS 2155$-$304
have accurate \ion{H}{1}\ column density measurements.  As a result, models 
of those HVCs are much better constrained.  For the comparisons of models to 
observations, we adopt 2$\sigma$ error bars for the measured column densities.
Allowing this flexibility in the comparisons is useful, given the simplicity of 
the models and the complexity of the actual situation.
In addition, in cases where line profiles indicate the presence
of saturation (\ion{C}{3} and \ion{Si}{3} for the $-140$ km s$^{-1}$ component
toward PKS 2155$-$304; \ion{Si}{3} for the
$-300$ km s$^{-1}$ component toward Mrk 509), we adopt the measured value 
as a lower limit to the column density.   The mean of the 
column densities obtained from the \ion{C}{4}\ doublet is used for the HVCs 
toward PKS 2155$-$304 since saturation is not present.  We have considered three
approaches for the ionization of these HVCs, which are discussed individually
below.  

\subsubsection{Photoionization by Galactic Starlight}

The first approach we consider is that the HVCs are in the Galactic halo
($d<50$ kpc) and photoionized by stars in the disk.  
For energetic reasons, UV photons from massive stars in the disk are 
thought to be responsible for ionizing a significant majority of the ionized
gas in the Galaxy and low halo (Reynolds 1993).  Some HVCs have been detected
in H$\alpha$ (Tufte et al.\ 2002), and the assumption of photoionization by
stars in the disk is often used to constrain their distances (Bland-Hawthorn \&
Maloney 2002).  In order to examine the measured column densities, we have 
generated a grid of photoionization models using the code CLOUDY 
(Ferland 1996).  We make the simplifying assumption that the absorbing gas can
be treated as plane-parallel slabs illuminated by incident radiation
dominated by OB associations.  The models assume a Kurucz model atmosphere 
($T_{\rm eff} = 35,000$~K) and include metallicity as a free parameter, 
although the relative abundance pattern is assumed to be solar.  Studies of 
H$\alpha$ emission measures of various HVCs have been argued (Bland-Hawthorn 
\& Putman 2001; Weiner, Vogel, \& Williams 2002) to be consistent with 
radiation from our Galaxy at the level of log $\phi\approx5.5$ (photons cm$^{-2}$ 
s$^{-1}$), where $\phi$ is the normally incident ionizing photon flux.
Since the intensity of the field is fixed, we run models for a range of 
different ionization parameter, $U$, by varying the hydrogen 
number density within the cloud ($U\propto\phi/n_{\rm H}$).

The CLOUDY models generated using a stellar ionizing spectrum are 
generally successful at predicting the measured column densities of 
the singly ionized and neutral species.  We find three generic failures
of these photoionization models: (1) All species doubly ionized and 
higher are underpredicted by models in which the low-ionization species
are used to constrain the field; (2) For any reasonable value of the 
ionization parameter, $U$, the stellar photoionization models produce very 
low ionization fractions of \ion{C}{4}\ relative to \ion{C}{2}; 
(3) These photoionization models are unable to produce the observed 
amount of \ion{O}{6}.  Clearly, if the gas is photoionized, 
the radiation field must be harder than that produced by massive stars
(the \ion{O}{5} ionization potential is $\chi_{\rm OV} = 113.87$ eV or 8.4 ryd). 

\subsubsection{Photoionization by Extragalactic Background}

Since a hard spectrum ionizing source is required to explain the ionization
pattern in these HVCs, a likely source is the integrated radiation from 
background QSOs and AGN.  This approach assumes, in effect, that the HVCs are 
at large distances from the Galactic plane ($d>100$ kpc) so that contributions 
from OB-star radiation escaping the galaxy are minimal (Giroux \& Shull 1997).
We use the same CLOUDY method as above, though we adopt a typical extragalactic 
radiation field with log $\phi=4.0$ (photons cm$^{-2}$ s$^{-1}$) and a power-law 
spectrum with a spectral index of $\alpha=-1.8$ (Zheng et al.\ 1997; Telfer et 
al.\ 2002). 

However, we find that single-parameter photoionization models cannot
reproduce the HVC ionization pattern.  For the stronger HVC
components, we observe $N$(\ion{C}{2})$\sim N$(\ion{C}{3})$\sim
N$(\ion{C}{4})$\sim N$(\ion{O}{6}), within a factor of 3.  No model
with our prescribed AGN radiation field can simultaneously predict
such an ionization pattern.  If we constrain the models by
$N$(\ion{C}{2})$\sim$$N$(\ion{C}{4}), then $N$(\ion{O}{6}) is
underpredicted by several orders of magnitude.  When
$N$(\ion{C}{2})$\sim$$N$(\ion{O}{6}) is assumed, then $N$(\ion{C}{4})
and $N$(\ion{N}{5}) are overpredicted by at least an order of
magnitude.  Although the weaker HVC components do not have measurable
amounts of low-ionization species, the ionization pattern is still
inconsistent with an extragalactic QSO field characterized by a single
value of $U$.  Telfer et al.\ (2002) show that quasar spectral indices
in the ionizing EUV (1--3 ryd) can vary widely, from $\alpha = 0$ to
$-3$.  In some cases, the spectrum can be considerably harder than our
adopted $\alpha=-1.8$ ionizing spectrum.  We explored the possibility
of a harder field by adopting $\alpha=-0.5$, which was also inadequate
in replicating the observations.  Although such a spectrum is able to
produce more \ion{O}{6} (8.4 ryd radiation), it does so at the expense
of lower ions which are still observed at high column density.
Additionally, we have explored models incorporating an extragalactic
X-ray background.  Although such a radiation field produces a larger
\ion{O}{6}\ column density by a factor of $3-5$, it is insufficient to
reproduce the observations.

Though they did not observe \ion{O}{6}, S99 were able to explain
simultaneously all the observed column densities within their
2$\sigma$ error bars by adopting a similar QSO-type field.  However,
their models predict values of $N$(\ion{O}{6}) significantly lower
than we observe, suggesting that \ion{O}{6} may be predominantly
collisionally ionized.  Since these models clearly have difficulty
explaining the measured \ion{O}{6} columns, we next consider models
which fit the set of column densities for species ionized to
\ion{N}{5} and below, as per S99.  We consider the species ionized
below \ion{O}{6}\ as arising from photoionization by the extragalactic
background, while the \ion{O}{6}\ arises from collisional ionization.

In Figures 9 and 10, we plot the predicted column densities versus ionization
parameter from the models of the $-140$ km s$^{-1}$ and 
$-270$ km s$^{-1}$ components, respectively, toward PKS 2155$-$304. 
The solid lines connecting the model data points define the 2$\sigma$
error bars of the column density measurements.  Neutral species are not 
plotted, since their upper limits do not provide 
useful constraints on the field.
Except for \ion{O}{6}, the observed column
densities can be simultaneously fitted by a QSO-type extragalactic field.    
The ionization parameters of the models fitting the $-140$ km s$^{-1}$ and 
$-270$ km s$^{-1}$ components are log $U=-3.02$ 
($n_{\rm H}=3.5\times10^{-4}$ cm$^{-3}$) and
log $U=-1.98$ ($n_{\rm H}=3.2\times10^{-5}$ cm$^{-3}$), 
respectively.  The \ion{H}{1}
column densities of the Mrk 509 HVC components are not well constrained, 
although adequate fits to species ionized below \ion{O}{6}\ can be obtained 
with similar metallicities, \ion{H}{1} column densities, and ionization 
parameters.  The fitted models for all four HVC components  
in the PKS 2155$-$304 and Mrk 509 sight lines, however, 
underpredict $N$(\ion{O}{6}) by as much as several orders of magnitude.   
In \S\ 4.3.3 we explore models in which the \ion{O}{6} column densities arise
from collisional ionization.

As noted by S99, there are scaling laws that apply for optically thin
clouds (log $N$(\ion{H}{1})$<17.2$) with sub-solar metallicities.  The most 
important of these scaling laws is the column density relation for species
X, $N$(X) $\propto Z_{\rm X} N(\mbox{\ion{H}{1}})$, where $Z_{\rm X}$
is the metallicity of element X.  Therefore, models with the same value
of $Z_{\rm X} N(\mbox{\ion{H}{1}})$ predict essentially the same 
runs of column density versus $U$.  The HVC components toward PKS 2155$-$304 
have well-constrained \ion{H}{1}\ column densities.  Since the fits to the 
observations are valid over a fairly narrow range in $U$, the metallicity and 
other properties can then be constrained by 
the uncertainty in $N$(\ion{H}{1}) and the use of the column density 
scaling law.  In accordance with that scaling law, 
we have performed model calculations for the PKS 2155$-$304
HVC components over the range of uncertainty in $N$(\ion{H}{1}).
Since only weak upper limits, log N(\ion{H}{1}) $< 17.69$, can be
derived for the HVCs toward Mrk 509, the parameter space of photoionization 
models is not well constrained.  Therefore, we restrict the following discussion 
on cloud properties and physical conditions to the PKS 2155$-$304 sight line.

\placetable{t6}

We obtain metallicities for the $-140$ km s$^{-1}$ and $-270$ km s$^{-1}$ 
components toward PKS 2155$-$304 of $\log Z_{140} = -0.47^{+0.15}_{-0.24}$ and 
$\log Z_{270} = -1.20^{+0.28}_{-0.45}$, respectively, similar to those in
HVC Complex C (CSG).  For the photoionized component of the HVCs, we can 
compute the pressure, $P/k=2.3n_{\rm H}T$ (assuming a fully ionized gas with 
He/H$=0.1$), and the cloud size, $D \propto N_{\rm H}/n_{\rm H}$.  
A summary of the properties obtained from the models is shown in Table 6. 
Shown for each component are three models using \ion{H}{1}\ columns constrained 
by the observations.  The inferred pressures, $P/k\sim1-10$ K cm$^{-3}$, 
are somewhat low, consistent with the S99 results for the HVCs toward Mrk 509.
Such a pressure is significantly below the expected coronal gas
pressure, $P/k\gtrsim100$ K cm$^{-3}$, in the Galactic halo.  Assuming that
the HVCs are in pressure equilibrium would imply distances well beyond the 
Galactic halo. The inferred size, $D \sim N_H / n_H$, of the $-270$ km s$^{-1}$ 
component, $D_{270} = 54^{+86}_{-24}$ kpc, is an order of magnitude larger than 
that of the $-140$ km s$^{-1}$ component, $D_{140} = 4.0^{+2.9}_{-1.4}$ kpc.   
This fact suggests an intergalactic location for the $-270$ km~s$^{-1}$ HVC.

However, we note that the ionization parameter, and thus the gas pressure, 
in these models is constrained primarily by the observed \ion{C}{4} and
\ion{Si}{4} column densities.  To explore the role of high-velocity 
\ion{C}{4} and \ion{Si}{4} detections on photoionization modeling, we 
have considered CLOUDY models of Complex C along the sight line toward 
PG~1259+593, irradiated by the same QSO-type field with
$\alpha=-1.8$.  Complex C is a large ensemble of HVCs 
believed to be falling into the Galaxy from the low halo (Wakker et al.\
1999; CSG).  However, significant columns of 
\ion{Si}{4}, \ion{C}{4}, and \ion{O}{6} associated with Complex C
are detected in this sight line.  As stated in the previous section, a 
radiation field arising from Galactic starlight cannot produce significant 
columns of \ion{C}{4} and \ion{Si}{4}.  If we constrain the ionization parameter 
of a QSO-type field by the \ion{C}{4} and \ion{Si}{4} columns observed towards 
PG~1259+593, we obtain the same low densities and low pressures as those
found in the models for the PKS 2155$-$304 HVCs.  However, the observed 
column densities, log~N(H~I) $\approx 19-20$, and transverse sizes, 100-300 pc,
for the Complex-C gas (CSG) suggest $n_H \sim 0.01-0.1$ cm$^{-3}$, in 
contradiction to the low densities needed to achieve a high $U$ parameter.  
The models in CSG (Figure 16) also suggest such a density.  Similar densities
are obtained by 
Wakker et al. (1999), who find that $n_H \sim 0.05-0.1$ cm$^{-3}$ for 
Complex C, 
based on a $5-25$ kpc range of possible distance.

Therefore, we find that the PKS 2155$-$304 HVCs have similar physical 
conditions to those in Complex C, but at much lower $N$(\ion{H}{1}).  
Since Complex C is a Galactic halo object, it is not unreasonable to assume 
that the PKS 2155$-$304 HVCs are as well.  
We further suggest that neither the physics of the highly ionized gas nor
its connection to the neutral and low ionization gas are adequately understood.

\subsubsection{Contributions from Collisional Ionization}

We believe that the strong \ion{O}{6} absorption in these ionized HVCs 
requires a significant contribution from collisional ionization. 
As demonstrated above, the observed column densities of species ionized
below \ion{O}{6}\ in these HVCs can be explained by photoionization from 
an extragalactic background produced by QSOs and AGN.  In the absence 
of an unusually hard ionizing field, gas at $T\sim10^{5.5\pm0.3}$~K is 
required to produce significant amounts of \ion{O}{6}.  Such gas 
probably arises in cooling layers, over a range of temperatures at
$T < 10^6$~K.  

Although a single-temperature collisional model cannot 
explain all the observed column densities, it is often useful to explore 
the ionization ratios using ionization fractions for collisional 
ionization equilibrium at fixed $T$ (CIE; Sutherland \& Dopita 1993).
We have therefore calculated the expected column densities based on 
$N$(\ion{H}{1}) in the collisionally ionized regions.  Since it is 
primarily the \ion{O}{6}\ that is underpredicted by the extragalactic 
photoionization models, we take the approach of considering
that it arises in a separate, collisionally ionized component.  
Such a component could manifest itself as a mixing boundary between the 
photoionized cloud core and the WHIM. With a solar abundance pattern,
this absorber would produce virtually no low-ionization species, 
while producing $2-3$ orders of magnitude lower \ion{C}{4} and \ion{N}{5} 
column densities than \ion{O}{6}.  (Gas that cools from $10^6$~K to $10^4$~K 
would produce additional \ion{C}{4}.)   Assuming that the 
metallicities in the photoionized and collisionally ionized regions are
equal, we can then use the measured values
of $N$(\ion{O}{6}) to calculate $N$(H) and $N$(\ion{H}{1}) in the 
collisionally ionized regions.  The contribution of photoionized gas 
to $N$(H) is given by the photoionization models.  

The $N$(\ion{O}{6}) observations imply nearly as 
much $10^{5.5}$ K collisionally ionized gas as photoionized gas, while 
contributing 2 to 4 orders of magnitude less in $N$(\ion{H}{1}) than the 
photoionized gas.  To explain the observations, the 
$-140$ km s$^{-1}$ component requires a factor of 3 less collisionally 
ionized than photoionized gas, although the $-270$ km s$^{-1}$
component would have nearly equal amounts of each.
However, if the collisionally ionized component is in pressure equilibrium 
with the photoionized component, then the implied size of the collisionally 
ionized component is more than an order of magnitude larger than the 
photoionized component.  Reflecting on the photoionized cloud sizes in
Table 6, we believe this model is unreasonable. One possible way of 
explaining this discrepancy would be if the collisionally ionized component 
was sufficiently dense to allow radiative cooling.  The cooling time of 
low-density hot gas is approximately 
\begin{equation}
   t_{\rm cool} = \frac {1.5 k T} {n_H \Lambda(T)} \approx
           \frac {(2.1~{\rm Gyr}) T_{6}} {n_{-4} \Lambda_{-22.5}} \; ,
\end{equation}
where $n_{-4} = n_H/(10^{-4}~{\rm cm}^{-3})$ and $T_{6} = (T/10^{6}$~K).
Here, $\Lambda_{-22.5}$ is the radiative cooling rate coefficient in units of
$10^{-22.5}$ erg~cm$^3$~s$^{-1}$, typical of gas at $10^{6}$~K
with 0.1 -- 0.3 solar metallicity (Sutherland \& Dopita 1993). 
At low densities, these cooling layers may be out of ionization equilibrium.    
Some or all of the observed \ion{C}{4} and \ion{Si}{4} could be attributed 
to the collisionally ionized component, thus allowing 
for lower ionization parameters and smaller cloud sizes.  The possible 
multi-component structure seen in Figure 2 for the $-140$ km s$^{-1}$ 
component towards PKS 2155$-$304, is suggestive
of such an association between the \ion{C}{4}, \ion{Si}{4}, \ion{O}{6}, and
possibly the doubly ionized components.  Indebetouw \& Shull (2003) summarize 
the predictions of a variety of theoretical models of interactions between 
gas in two thermal phases, such as evaporating cloudlets (Ballet et al.\ 1986),
planar conduction fronts (Borkowski et al.\ 1990), stellar wind 
and supernovae bubbles (Slavin \& Cox 1993), turbulent mixing layers
(Slavin, Shull, \& Begelman 1993), and more generic cooling flows. 

A specific scenario consistent with the ionization and kinematic 
features of the \ion{O}{6} is the ``infalling clump" model. In this
picture, the HVCs represent clumps of gas, falling into the
Galactic halo after large distances or returning to the disk from
injection into a Galactic fountain.  (The metallicity of the gas would 
be a distinguishing feature for these two scenarios.)  Interactions
between the clump and the gaseous substrate of the low halo would
produce a bow shock, and the refraction and shear of the post-shock
gas will create a range of observed radial velocities, of order
$(1-2) V_s \sin b$, if the HVC is falling directly toward the plane
at Galactic latitude $b$ and velocity $V_s$.  Such velocity shear 
has been seen in the bow shocks toward many Herbig-Haro objects
(Hartigan, Raymond, \& Hartmann 1987), and may be reflected in the 
broad ($\sim 100$ km~s$^{-1}$) profiles of \ion{O}{6}.      

We can also consider more generic predictions of the \ion{C}{4} and 
\ion{Si}{4} which may be associated with these nonequilibrium physical states.  
In general, substantially more \ion{C}{4} and \ion{Si}{4} may co-exist 
with the \ion{O}{6} than would be inferred from collisional ionization 
equilibrium. While the fractions of N(\ion{C}{4})/N(\ion{O}{6}) and 
N(\ion{Si}{4})/N(\ion{O}{6}) are ultimately dependent on the details
of the specific model, a representative range in these fractions is 
N(\ion{C}{4})/N(\ion{O}{6}) $\approx 0.1-0.2$
and N(\ion{Si}{4})/N(\ion{O}{6}) $\lesssim0.01$.  If we associate
10-20\% of our measured column densities of N(\ion{C}{4})
and N(\ion{Si}{4}) with this hot gas, this is not enough
to change the inferred physical properties of the
photoionized gas substantially.

We can make a substantial change in the acceptable ionization parameter
for the photoionized gas, and thus in the inferred density, if nearly all 
the multiply ionized species (\ion{O}{6}, \ion{C}{3}, \ion{C}{4}, 
\ion{Si}{3}, \ion{Si}{4}, \ion{N}{5}) are associated with collisionally 
ionized gas.  If the radiation field from the CLOUDY models is then 
constrained by \ion{H}{1} and singly-ionized species, significantly lower 
photoionization parameters are possible.
As a result, the $-140$ km s$^{-1}$ component could have a gas density 
as much as six times larger than the values listed in Table 6, and 
the gas density of the $-270$ km s$^{-1}$ component could be larger by an 
even greater factor. In such a scenario, a larger gas pressure 
would make an association with the Galaxy more likely for these HVCs.  

\subsubsection{Possible contributions from the WHIM}

X-ray absorption lines of \ion{O}{7} and \ion{O}{8} are 
detected at $z=0$ in the PKS 2155$-$304 sight line, both with {\it Chandra} 
(Nicastro et al. 2002) and XMM (Rasmussen, Kahn, \& Paerels 2003).  
In this section we consider the implications of these detections on the 
characteristics of the absorbing gas.  Since the velocity resolution of the
X-ray data is inferior to that of the UV data by a factor of $\sim30$, the
\ion{O}{7} and \ion{O}{8} features cannot be conclusively associated with
the high-velocity UV lines.  Since \ion{O}{6} components at $z=0$ are detected
in this sight line, both at high velocity and Galactic LSR velocity, it
is likely that the X-ray absorption features also have multi-component
structure, albeit unresolved.  Nevertheless, we consider the scenario where 
the entire \ion{O}{7} and \ion{O}{8} absorption arises in a single component 
at high-velocity.

We adopt log[$N$(\ion{O}{8})/$N$(\ion{O}{7})]$=0.12^{+0.34}_{-0.69}$ based on 
the column densities reported in Nicastro et al. (2002).  
This corresponds to a 
temperature of log~$T=6.30\pm0.15$, assuming collisional ionization equilibrium
(Sutherland \& Dopita 1993).  The \ion{O}{6} to \ion{O}{7} ionization 
ratio in CIE would then be log[$f$(\ion{O}{6})/$f$(\ion{O}{7})$]=-2.3$. 
Therefore, if the entire \ion{O}{7}/\ion{O}{8} absorption occurs in
a WHIM component at the same velocity as the high-velocity UV absorption, 
with $N$(\ion{O}{7})$=4.0 \times 10^{15}$ cm$^{-2}$, then
the WHIM could account for a column density, $N$(\ion{O}{6})$=2.0\times10^{13}$
cm$^{-2}$.  Since some of the O~VII/VIII absorption could be attributed to a 
Galactic component at $V_{LSR}\approx0$ km s$^{-1}$, we stress that this value
is an upper limit to the amount of \ion{O}{6} that would exist in the
high-velocity WHIM.  In the two high-velocity components towards PKS
2155$-$304, 
we measure a total column density of $N$(\ion{O}{6})$=1.0\times10^{14}$
cm$^{-2}$, 
a factor of five larger than could be accounted for by a velocity-coincident
WHIM absorber.  It is also of note that the \ion{O}{6} absorption is
confined to a velocity range of half-width $\sim100$ km s$^{-1}$, whereas the 
shocked WHIM is expected to extend over a broader region, at distances of
$2-25 h^{-1}$ Mpc, or at $V_{LSR}=200-2500$ km s$^{-1}$
(Kravtsov, Klypin, \& Hoffman 2002).  
We therefore conclude that most of the \ion{O}{6} gas is 
present in the high-velocity cloud traced by C~II/III/IV and Si~II/III/IV.
We emphasize that this does not rule out the possibility that the O~VII/VIII 
absorber may trace a WHIM filament, or that some of the O~VI absorption arises
in the WHIM.

\section{CONCLUSIONS}

Using {\it FUSE} and {\it HST}/STIS data for the targets PKS 2155$-$304 and
Mrk~509, we have investigated the ``highly ionized high-velocity
clouds" in those sight lines.  We measured column densities of species 
associated with the HVCs, as well as the \ion{H}{1} column
density of the HVCs toward PKS 2155$-$304 using Lyman-series absorption.
Using these results, we have come to the following conclusions:

\noindent
{\bf (1) Highly Ionized HVCs.}   
With {\it HST} and {\it FUSE}, we have studied four ``highly ionized" HVCs 
proposed by S99, observing \ion{H}{1} and multiple ion stages of carbon, 
silicon, and \ion{O}{6}.  We confirm the ionized nature of three of the four 
HVCs, the exception being the HVC at $-140$ km~s$^{-1}$ 
toward PKS~2155$-$304, which has many low ions (\ion{H}{1}, \ion{C}{2}, 
\ion{Si}{2}).  A core-halo effect is apparent with high-ionization species 
(doubly-ionized and above) which show a larger extent in velocity space 
than the low-ionization species.  Such behavior may indicate a difference 
in physical origin between the high- and low-ionization gas.

\noindent
{\bf (2) Multiphase Gas.}  
We observe far more \ion{O}{6} in these HVCs than is consistent with 
photoionization models; we believe \ion{O}{6} is collisionally ionized. 
The HVCs exhibit absorption ranging from \ion{H}{1} to \ion{C}{4}, 
\ion{Si}{4}, and \ion{O}{6}, suggesting that the clouds are multiphase.  
The combination of hot collisionally ionized gas with warm
photoionized gas complicates the interpretation.   
We propose that the high ions arise from hot gas, possibly produced 
behind bowshocks from clumps falling toward the Galactic plane.  
Velocity shear behind these bowshocks is consistent with the observed 
broad ($\sim100$ km~s$^{-1}$) \ion{O}{6} absorption profiles.  

\noindent
{\bf (3) Column Density Ratios.}   Some of the
ion ratios in these HVCs differ from 
most Galactic halo gas.  The ionization pattern in the stronger HVC components
has $N$(\ion{C}{2})$\sim N$(\ion{C}{3})$\sim N$(\ion{C}{4})$\sim N$(\ion{O}{6})
within a factor of 3.   Some of these ratios are similar to intergalactic 
gas, both the high-$z$ Ly$\alpha$ forest and a low-$z$ absorber associated with 
the Virgo cluster.  However, the presence of low ions indicates that the density 
is significantly higher than that expected for the WHIM.  
This implies, at least for the HVCs towards PKS 2155$-$304 and Mrk 509, 
that the high-velocity \ion{O}{6} do not necessarily trace the WHIM,
as proposed by Nicastro et al.\ (2002, 2003), but
instead may trace HVCs with low total hydrogen column density.  
Some \ion{C}{4} and \ion{Si}{4} could arise from thermal instability in 
cooling WHIM gas, but this may be inconsistent with the broad \ion{O}{6} 
velocity profiles.

\noindent
{\bf (4) Photoionization models.}  Single-parameter ($U$) models cannot 
reproduce all the observed column densities. If \ion{O}{6} is assumed to be 
collisionally ionized, a QSO-type radiation field can 
reproduce the other species' column density pattern.  In this case, the 
radiation field is constrained by \ion{C}{4} and \ion{Si}{4}, which imply 
low gas pressures, large cloud sizes, and an extragalactic location for the 
HVCs.  However, models of the PG~1259+593 sight line toward HVC Complex C 
(thought to have $n_H \sim 0.01-0.1$ cm$^{-3}$) yield the same, erroneously low
gas pressures when the field is constrained by the high-ion measurements.  
This fact indicates that the physics of hot gas at cloud interfaces may not be 
well understood.    

\noindent
{\bf (5) HVC Gas Pressures.}  
Simple cooling layers, from $10^6$~K down to $10^4$~K, cannot account for the 
\ion{C}{4} and \ion{Si}{4} measurements, given the strong absorption in 
\ion{O}{6}.  A substantial decrease in the ionization parameter, $U$,
and increase in the gas pressure, $P/k$, is possible in a multiphase model
in which species doubly-ionized and above arise from collisional ionization.  
If photo-ionization models are then constrained by \ion{H}{1} and the
singly-ionized species, significantly larger gas pressures are allowed, and 
a Galactic association for the HVCs is possible.

\acknowledgments

This work was supported by grant GO-08571.01-A (HST Cycle 8) from the Space 
Telescope Science Institute and NAG5-7262 from NASA/LTSA.  The FUSE data were 
obtained by the Guaranteed Time Team for the NASA-CNES-CSA mission operated by 
the Johns Hopkins University.  Financial support to U.S. participants was provided 
by NASA contract NAS5-32985. We thank Ken Sembach and Blair Savage for 
for helpful discussions on Galactic halo gas and HVCs.  We also
thank the referee, Fabrizio Nicastro, for useful comments on the final
preparation of this paper.

\begin{figure}
\figurenum{1}
\plotone{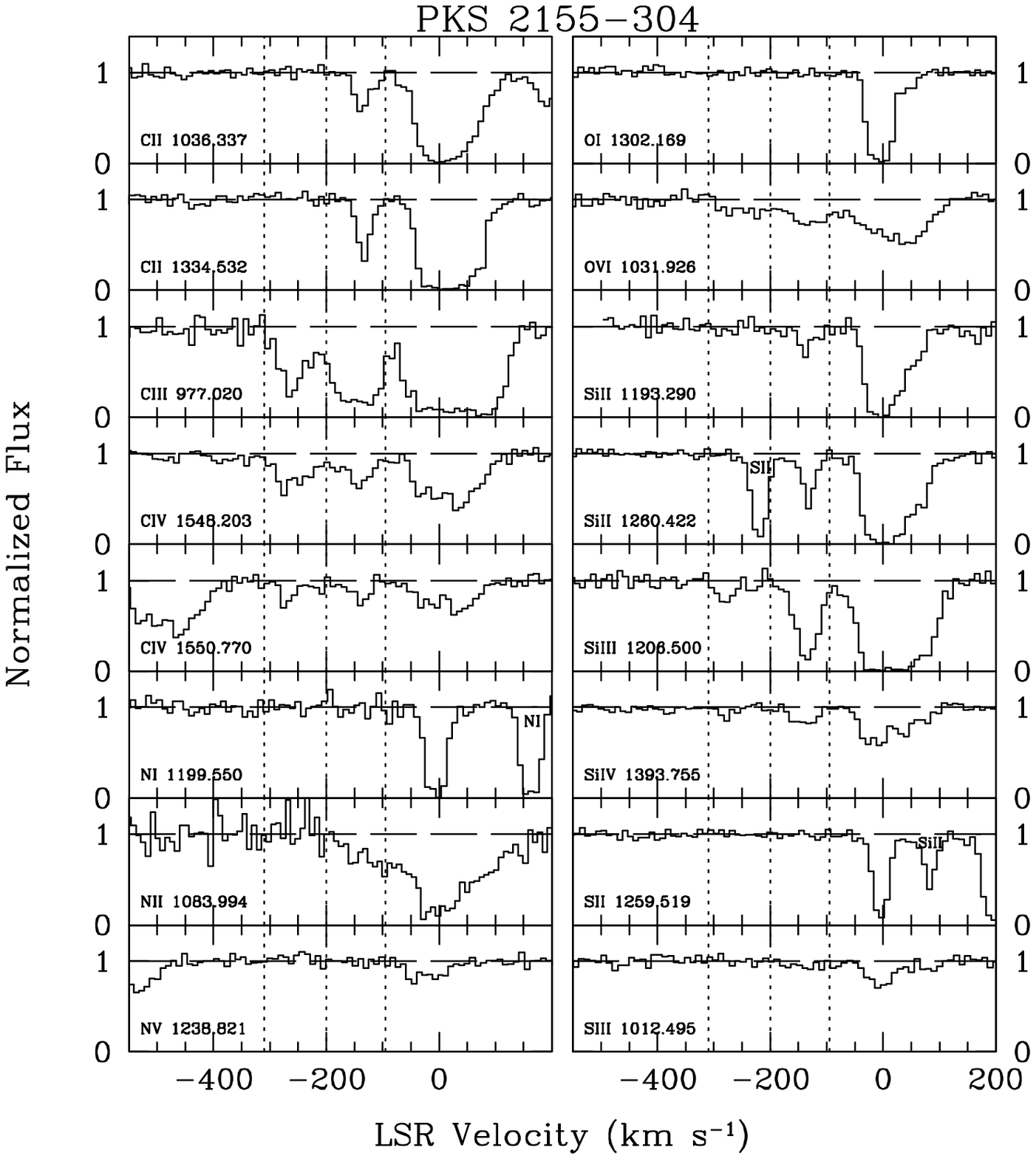}
\caption{Normalized absorption profiles from STIS and {\it FUSE} data for
PKS 2155$-$304.  The vertical dashed lines indicate the integration ranges
of  the $-140$ km s$^{-1}$ ($-200$ to $-95$ km s$^{-1}$) and the $-270$ 
km s$^{-1}$ ($-310$ to $-200$ km s$^{-1}$) components.}
\end{figure}

\begin{figure}
\figurenum{2}
\plotone{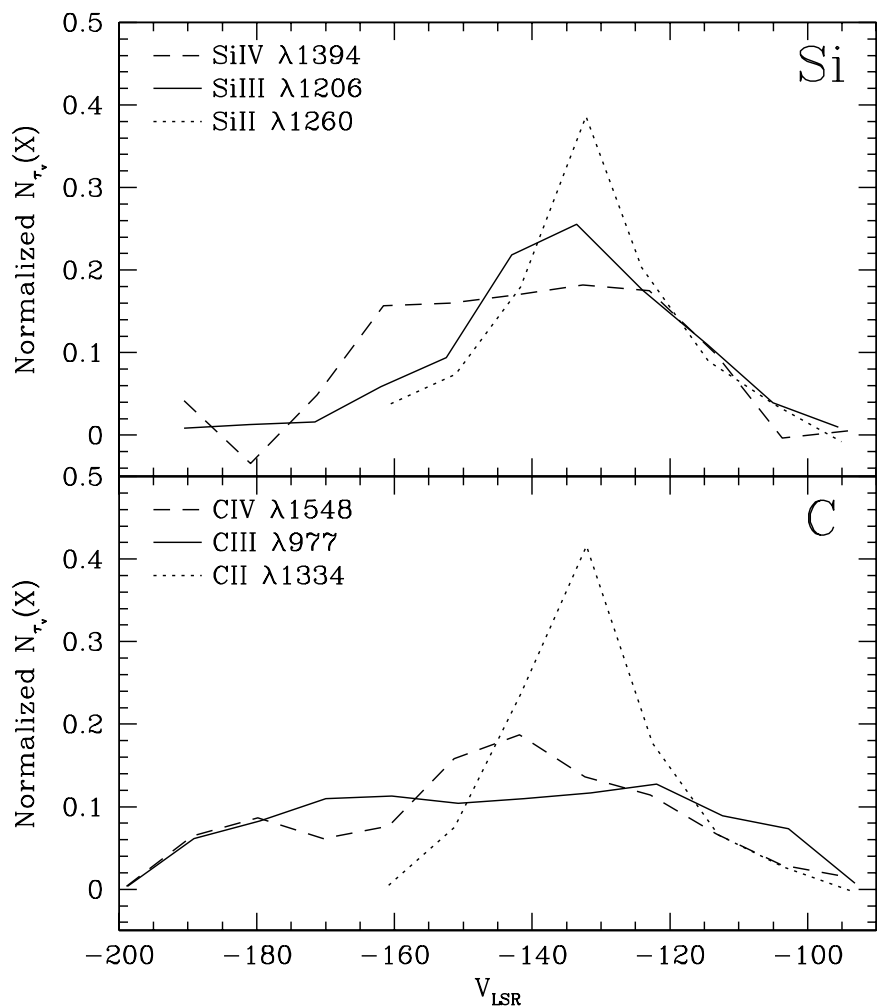}
\caption{Normalized column density profiles 
of silicon ({\it top}) and carbon ({\it bottom}) for the $-140$ km s$^{-1}$ 
component in the PKS 2155$-$304 sight line.  The more concentrated profile 
of the singly-ionized species indicates a possible core-halo structure
of the HVC.}   
\end{figure}

\begin{figure}
\figurenum{3}
\plotone{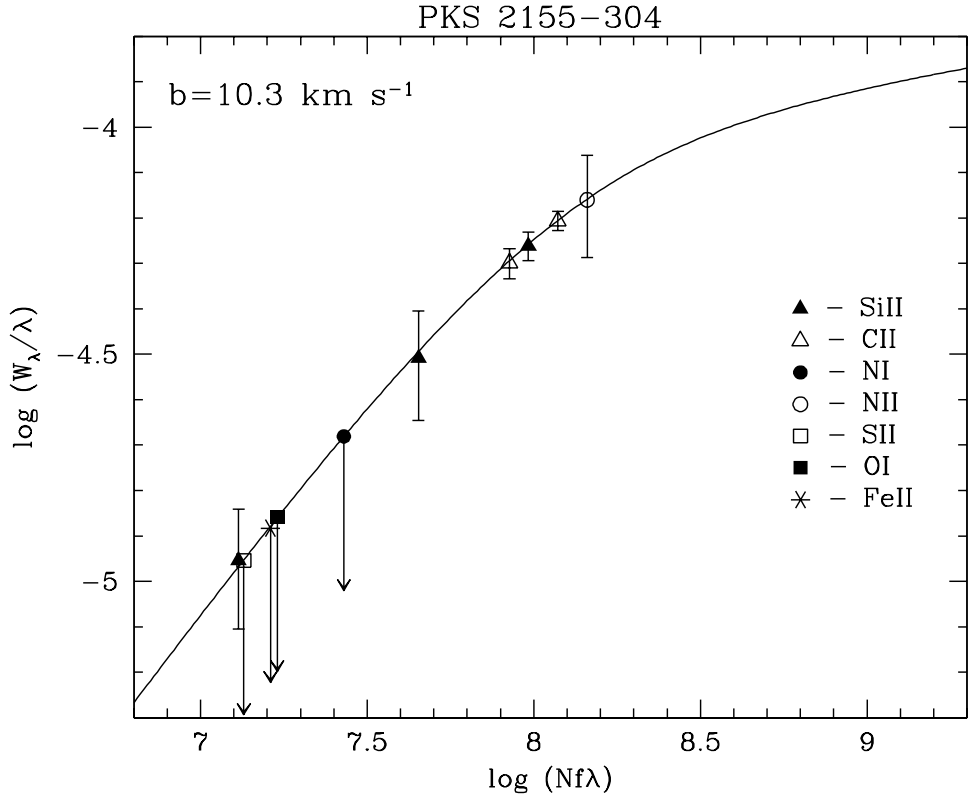}
\caption{Empirical curve of growth for ions observed in the PKS~2155$-$304 
    sight line ($-140$ km~s$^{-1}$ component). The \ion{Si}{2}\ and \ion{C}{2}\ 
    lines are best fitted by a curve with $b=10.3^{+6.5}_{-2.9}$ km s$^{-1}$.}
\end{figure} 

\begin{figure}
\figurenum{4}
\plotone{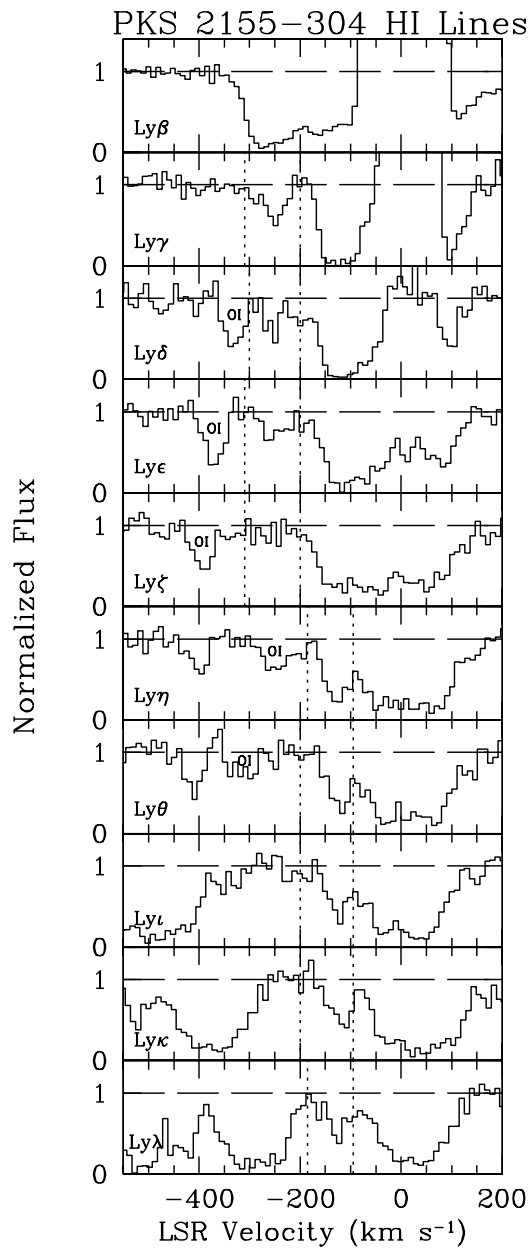}
\caption{Normalized absorption profiles of \ion{H}{1}\ Lyman 
lines from the PKS 
2155$-$304 {\it FUSE} data.  The vertical dashed lines indicate the Lyman lines
for which the $-140$ and $-270$ km s$^{-1}$ components can be measured, as
well as the adopted integration range.  The $-140$ km s$^{-1}$ component
can be measured for Ly$\gamma$ through Ly$\zeta$, while the $-270$ km s$^{-1}$
can be measured for Ly$\eta$ through Ly$\lambda$.} 
\end{figure} 

\begin{figure}
\figurenum{5}
\plotone{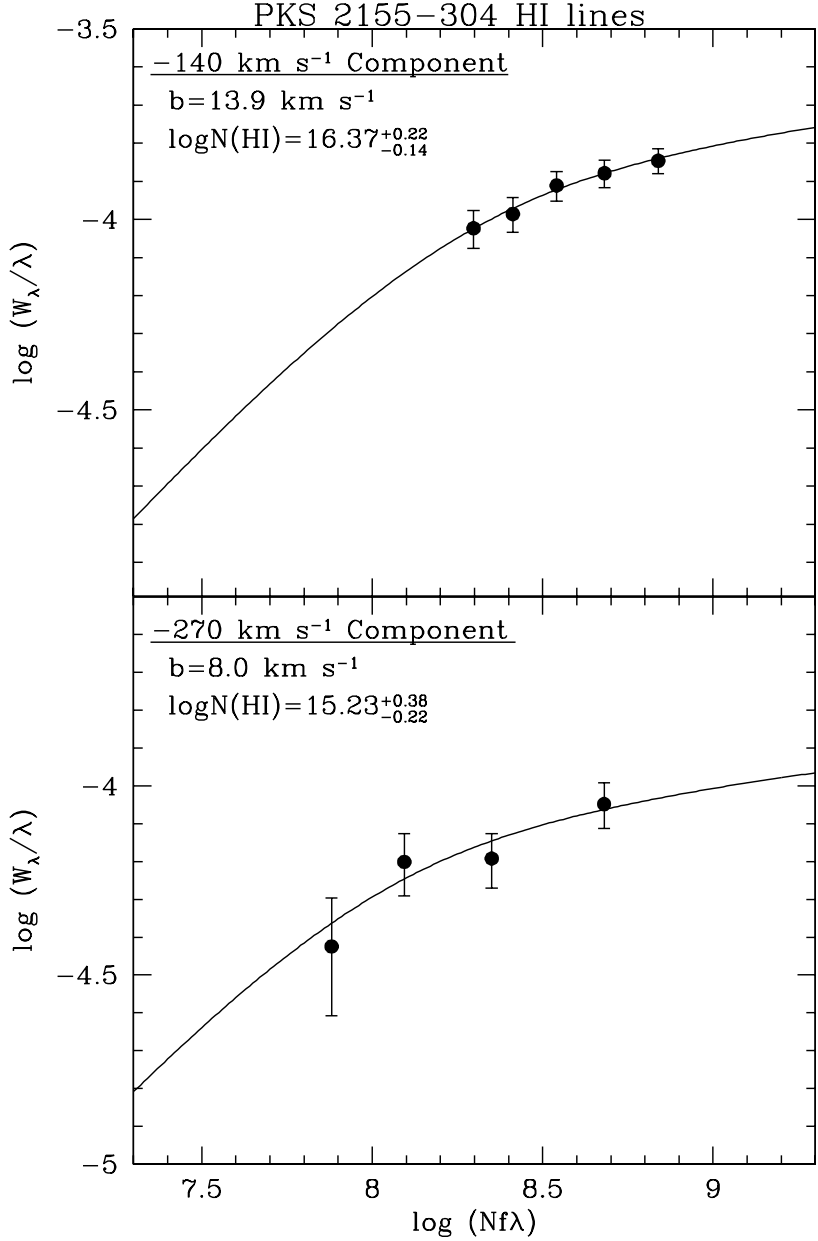}
\caption{Empirical curves of growth for \ion{H}{1} lines observed in the 
PKS~2155$-$304 sight line for the two high velocity components.  The $-140$ 
and $-270$ km s$^{-1}$ components are best fitted by curves with 
$b=13.9^{+3.1}_{-2.3}$ and $b=8.0^{+3.5}_{-2.0}$ km s$^{-1}$, respectively.  
The measured column densities are, respectively, 
log$N$(\ion{H}{1})$=16.37^{+0.22}_{-0.14}$ and 
log$N$(\ion{H}{1})$=15.23^{+0.38}_{-0.22}$.}
\end{figure} 

\begin{figure}
\figurenum{6}
\plotone{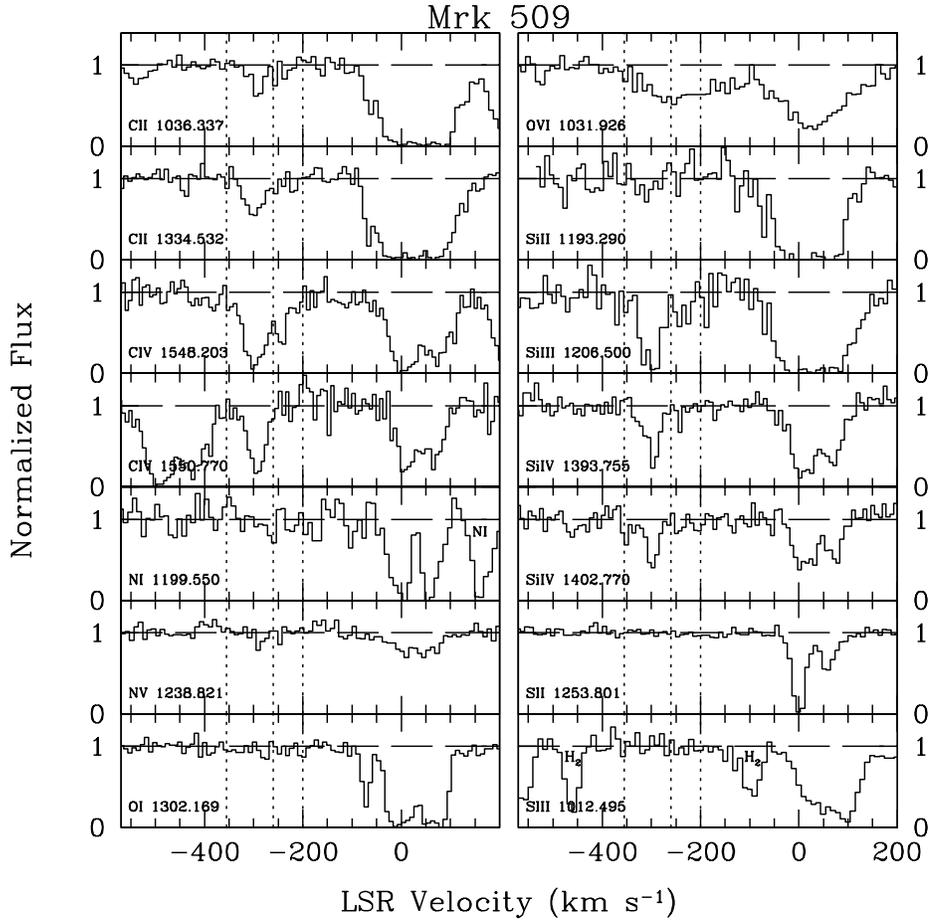}
\caption{Normalized absorption profiles from STIS and {\it FUSE} data for
Mrk 509.  The vertical dashed lines indicate the integration ranges
of  the $-240$ km s$^{-1}$ ($-260$ to $-200$ km s$^{-1}$) and the $-300$ 
km s$^{-1}$ ($-355$ to $-260$ km s$^{-1}$) components.}
\end{figure}

\begin{figure}
\figurenum{7}
\plotone{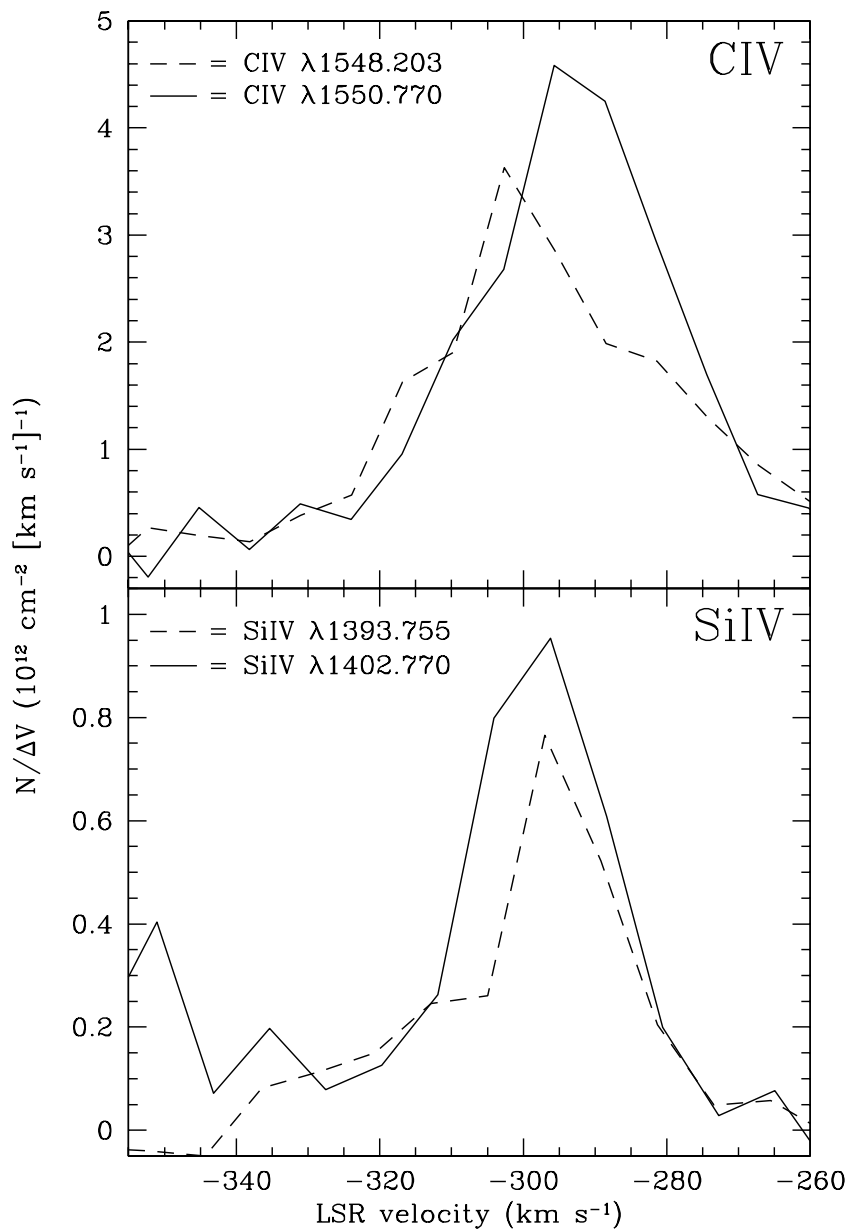}
\caption{Column density profiles of  \ion{C}{4}\ and \ion{Si}{4}\
for the $-300$ km s$^{-1}$
component in the Mrk 509 sight line.  Plotted are the profiles from the strong
({\it dashed}) and weak ({\it solid}) lines in each doublet.  The more peaked
behavior of the profiles from the weak lines indicates the presence of 
saturation.}
\end{figure}

\begin{figure}
\figurenum{8}
\plotone{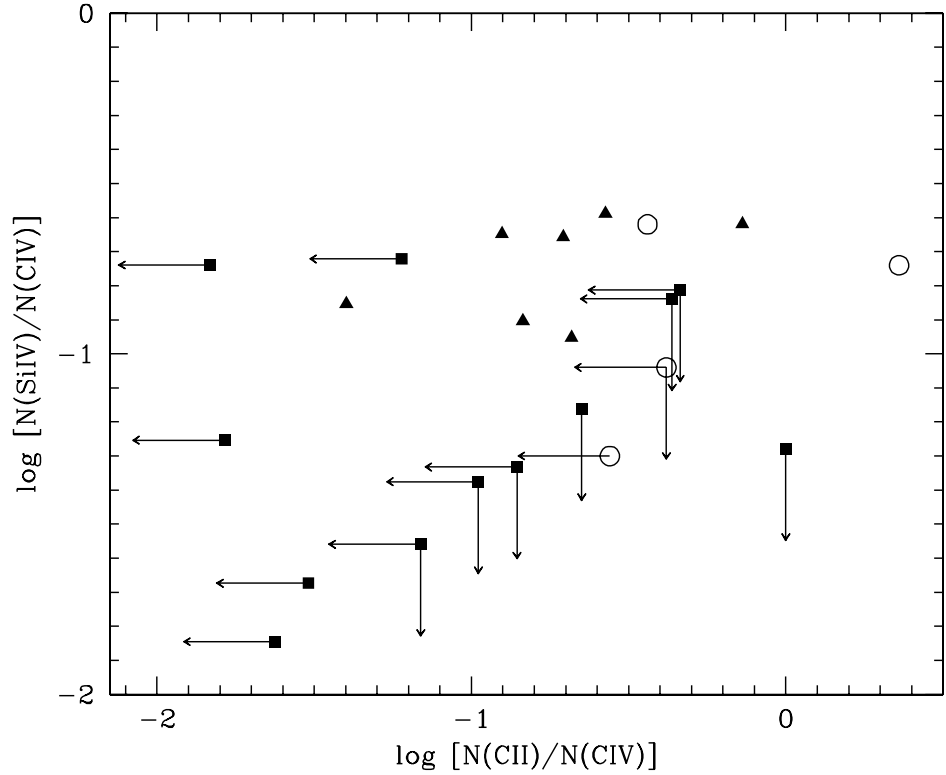}
\caption{Diagnostic diagram of the ratios log[$N$(\ion{Si}{4})/$N$(\ion{C}{4}] 
versus log[$N$(\ion{C}{2})/$N$(\ion{C}{4}] for $z\sim3$ Ly$\alpha$ forest 
clouds and the HVCs towards PKS 2155$-$304 and Mrk 509.  Triangles
are measurements and squares are limits from the Songaila \& Cowie (1996)
sample.  The large open circles are for the HVCs towards PKS 2155$-$304 and 
Mrk 509.  The data point with very high log[$N$(\ion{C}{2})/$N$(\ion{C}{4}] is
for the $-140$ km s$^{-1}$ component toward PKS 2155$-$304.}
\end{figure}

\begin{figure}
\figurenum{9}
\plotone{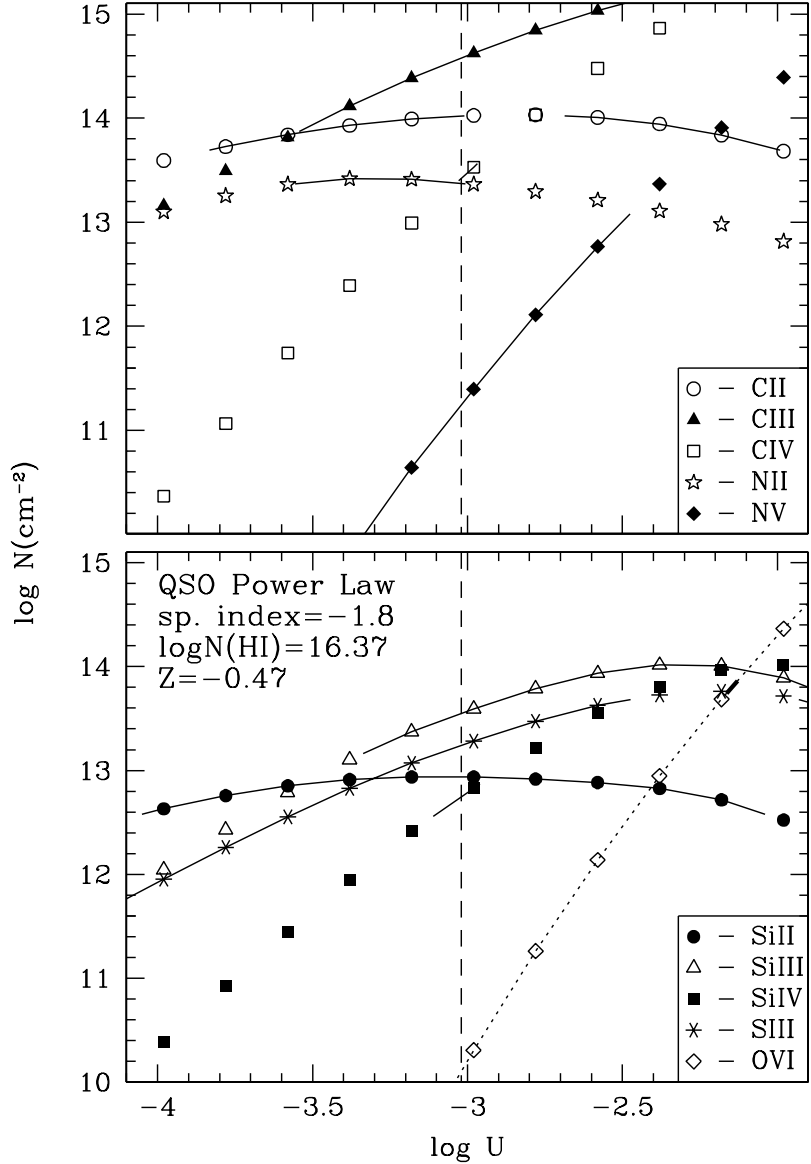}
\caption{CLOUDY model calculations for the $-140$ km s$^{-1}$ HVC component 
toward PKS 2155$-$304.  Plotted are the predicted column densities of various ion 
species versus log $U$ (points) 
and their constraints based on the observations (solid lines).  Species
are shown on two plots for clarity.
For the constraints on the observed values, we use the 2$\sigma$ range of 
the measured value, the measured value as a lower limit if saturation
is evident, or the upper limit on the measured value.  
The predicted run of $N$(\ion{O}{6}) is indicated by the dotted line to 
emphasize the discrepancy between the predicted and observed values of
$N$(\ion{O}{6}).  The model shown is 
characterized by a QSO power law spectrum with $\alpha=-1.8$, 
an \ion{H}{1} column density of log$N$(\ion{H}{1})$=16.37$ and a
metallicity of $Z = 10^{-0.47} Z_{\odot}$.  The vertical dashed line 
indicates a value of ionization parameter, log $U=-3.02$,
which fits the observed column densities, excluding \ion{O}{6}.  }
\end{figure}

\begin{figure}
\figurenum{10}
\plotone{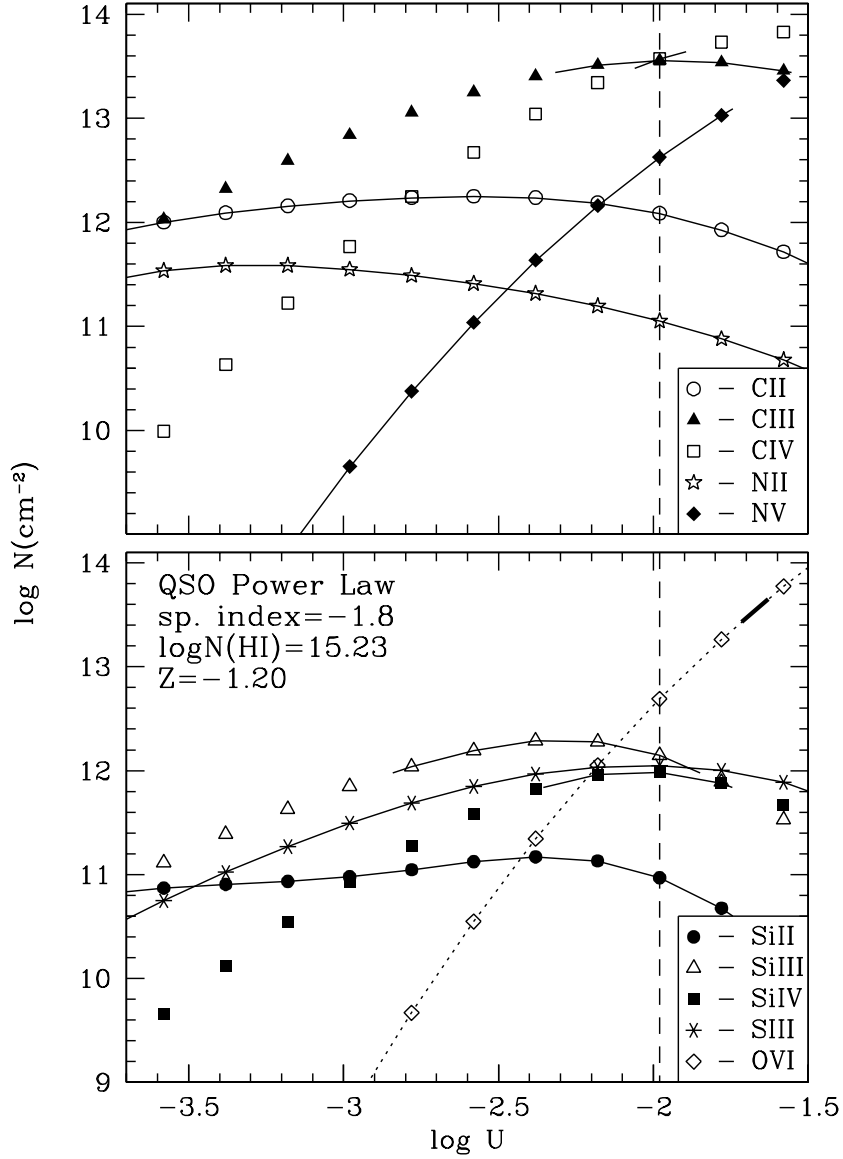}
\caption{CLOUDY model calculations for the $-270$ km s$^{-1}$ HVC component 
toward PKS 2155$-$304.  Plot symbols are the same as in Figure 8.  The model 
shown is characterized by a QSO power law spectrum with $\alpha=-1.8$, 
an \ion{H}{1} column density of log$N$(\ion{H}{1})$=15.23$ and a
metallicity of $Z = 10^{-1.20} Z_{\odot}$.  The vertical dashed line 
indicates a value of ionization parameter, log $U=-1.98$,
which fits the observed column densities, excluding \ion{O}{6}.}
\end{figure}

\clearpage
\begin{deluxetable}{llcclcc}
\tablecolumns{5}
\tablewidth{0pc}
\tablecaption{SUMMARY OF OBSERVATIONS \label{t1}}
\tablehead{
\colhead{} & \colhead{} & \colhead{} & \colhead{} & \colhead{} & \colhead{} & 
   \colhead{Number of} \\
\colhead{Sightline} & \colhead{Instrument} & \colhead{Grating} & 
   \colhead{Program ID} & 
   \colhead{Observation Date(s)} & \colhead{$T_{exp}$(ks)} & \colhead{Exposures} 
}
\startdata
PKS 2155$-$304 & {\it HST-STIS}   & E140M\tablenotemark{a} & 8125 & 1999 Nov;
      2000 Sep & 28.5  & 4  \\
    & {\it FUSE}\tablenotemark{b} & ... & P108 & 1999 Oct; 2001 Jun & 121.5 & 54 \\ 
Mrk 509  & {\it HST-STIS}   & E140M     & 8877 & 2001 Apr & 7.6 & 3  \\
         & {\it FUSE}       & ...       & P108 & 2000 Sep & 61.4  & 28 \\
\enddata
\tablenotetext{a}{The STIS E140M provides wavelength coverage of 
    1150--1700 \AA.} 
\tablenotetext{b}{The {\it FUSE} spectrum covers the wavelength range of 
    905-1187 \AA.}
\end{deluxetable}

\clearpage
\begin{deluxetable}{lrlccccc}
\tablecolumns{8}
\tablewidth{0pc}
\tablecaption{SUMMARY OF MEASUREMENTS: PKS 2155$-$304 \label{t2}}
\tablehead{
\colhead{} & \colhead{} & \colhead{} & \colhead{} & 
    \multicolumn{2}{c}{Component 1} & \multicolumn{2}{c}{Component 2} \\
\colhead{} & \colhead{} & \colhead{} & \colhead{} & 
   \multicolumn{2}{c}{$V_{LSR}=-140$ km s$^{-1}$} & 
   \multicolumn{2}{c}{$V_{LSR}=-270$ km s$^{-1}$} \\
\colhead{} & \colhead{$\lambda$\tablenotemark{a}} & \colhead{} & \colhead{} & 
   \colhead{$W_{\lambda}$} & \colhead{log $N(X)$\tablenotemark{b}} & 
   \colhead{$W_{\lambda}$} & \colhead{log $N(X)$\tablenotemark{b}} \\
\colhead{Species} & \colhead{(\AA)} & \colhead{$f$\tablenotemark{a}} & 
   \colhead{} & \colhead{(m\AA)} & \colhead{($N$ in cm$^{-2}$)} & 
   \colhead{(m\AA)} & \colhead{($N$ in cm$^{-2}$)}}
\startdata
\ion{H}{1}  & ...  \ \ & \ \ ...    & & ...   & 
   $16.37^{+0.22}_{-0.14}$\tablenotemark{c} & ...      & 
   $15.23^{+0.38}_{-0.22}$\tablenotemark{c}  \\
\ion{C}{2}  & 1036.337 \ \ & \ \ 0.118  & & $52\pm4$   & 
   $13.84^{+0.10}_{-0.07}$\tablenotemark{d} & $<18$    & $<13.21$ \\
   & 1334.532 \ \ & \ \ 0.128  & & $83\pm4$   & ...                            
          & $<20$      & $<13.00$                                  \\
\ion{C}{3}  &  977.020 \ \ & \ \ 0.759  & & $246\pm9$  & $13.87^{+0.02}_{-0.02}$*       
          & $149\pm11$ & $13.52^{+0.03}_{-0.04}$               \\
\ion{C}{4}  & 1548.203 \ \ & \ \ 0.190  & & $112\pm10$ & $13.51^{+0.03}_{-0.04}$        
          & $137\pm11$ & $13.60^{+0.04}_{-0.03}$                   \\
            & 1550.770 \ \ & \ \ 0.0948 & & $51\pm9$   & $13.45^{+0.07}_{-0.08}$        
          & $60\pm11$  & $13.52^{+0.07}_{-0.09}$                   \\
\ion{N}{1}  & 1199.550 \ \ & \ \ 0.130  & & $<25$      & $<13.24$\tablenotemark{d}      
          & $<35$      & $<13.32$                                  \\
\ion{N}{2}  & 1083.994 \ \ & \ \ 0.115  & & $75\pm19$  & 
   $14.06^{+0.22}_{-0.22}$\tablenotemark{d} & $<62$    & $<13.71$ \\
\ion{N}{5}  & 1238.821 \ \ & \ \ 0.157  & & $<25$      & $<13.08$                       
          & $<26$      & $<13.09$                                  \\
\ion{O}{1}  & 1302.169 \ \ & \ \ 0.0519 & & $<18$      & $<13.40$\tablenotemark{d}      
          & $<25$      & $<13.51$                                  \\
\ion{O}{6}  & 1031.926 \ \ & \ \ 0.133  & & $70\pm5$   & $13.80^{+0.03}_{-0.03}$        
          & $42\pm5$   & $13.56^{+0.05}_{-0.06}$                   \\
\ion{Si}{2} & 1193.290 \ \ & \ \ 0.585  & & $37\pm10$  & 
   $12.81^{+0.12}_{-0.10}$\tablenotemark{d} & $<42$    & $<12.77$ \\ 
            & 1260.422 \ \ & \ \ 1.184  & & $69\pm5$   & ...                            
            & ...        & ...                                  \\
            & 1526.707 \ \ & \ \ 0.132  & & $17\pm5$   & ...                            
          & $<34$      & $<13.10$                                  \\ 
\ion{Si}{3} & 1206.500 \ \ & \ \ 1.669  & & $176\pm11$ & $13.16^{+0.02}_{-0.02}$*       
          & $41\pm12$  & $12.33^{+0.10}_{-0.14}$                   \\
\ion{Si}{4} & 1393.755 \ \ & \ \ 0.514  & & $45\pm8$   & $12.74^{+0.07}_{-0.08}$        
          & $15\pm5$   & $12.26^{+0.11}_{-0.16}$                   \\
\ion{S}{2}  & 1259.519 \ \ & \ \ 0.0166 & & $<14$      & $<13.81$\tablenotemark{d}      
          & $<20$      & $<13.94$                                  \\
\ion{S}{3}  & 1012.495 \ \ & \ \ 0.0442 & & $<19$      & $<13.68$                       
          & $<20$      & $<13.69$                                  \\
\ion{Fe}{2} & 1144.938 \ \ & \ \ 0.106  & & $<15$      & $<13.12$\tablenotemark{d}      
          & $<21$      & $<13.23$                                  
\enddata
\tablecomments{* Because some saturation is indicated in the line profile, the
    total column density may be slightly higher.}
\tablenotetext{a}{Wavelengths and oscillator strengths are from D. C. Morton 
   (2003, in preparation).}
\tablenotetext{b}{Unless otherwise specified, column densities are calculated 
through the apparent optical depth method.  All upper limits are 3$\sigma$ levels.}
\tablenotetext{c}{\ion{H}{1}\ column densities are calculated through a 
   curve-of-growth analysis of Lyman series lines in the {\it FUSE} bandpass.}
\tablenotetext{d}{Column density is calculated through a curve-of-growth 
   analysis of low-ionization species.  The data is best fitted by a 
   curve of growth with doppler parameter, $b=10.3^{+6.5}_{-2.9}$ km s$^{-1}$.} 
\end{deluxetable}

\clearpage
\begin{deluxetable}{lrlccccc}
\tablecolumns{8}
\tablewidth{0pc}
\tablecaption{PKS 2155$-$304 \ion{H}{1} MEASUREMENTS \label{t3}}
\tablehead{
\colhead{} & \colhead{} & \colhead{} & 
   \multicolumn{2}{c}{Equivalent Width ($W_{\lambda}$ in m\AA)} \\
\colhead{} & \colhead{} & \colhead{} & \colhead{$-140$ km s$^{-1}$} & 
   \colhead{$-270$ km s$^{-1}$} \\
\colhead{Line} & \colhead{$\lambda$(\AA)} & \colhead{$f$\tablenotemark{a}} & 
   \colhead{Component\tablenotemark{b}} & \colhead{Component\tablenotemark{c}}}
\startdata
Ly$\gamma$   & 972.537 \ \ & \ \ 0.0290   & ...        & $87\pm12$   \\
Ly$\delta$   & 949.743 \ \ & \ \ 0.0139   & ...        & $61\pm10$   \\
Ly$\epsilon$ & 937.803 \ \ & \ \ 0.00780  & ...        & $59\pm11$   \\
Ly$\zeta$    & 930.748 \ \ & \ \ 0.00481  & ...        & $35\pm12$   \\
Ly$\eta$     & 926.226 \ \ & \ \ 0.00318  & $132\pm10$ & ...         \\
Ly$\theta$   & 923.150 \ \ & \ \ 0.00222  & $122\pm10$ & ...         \\
Ly$\iota$    & 920.963 \ \ & \ \ 0.00161  & $113\pm10$ & ...         \\
Ly$\kappa$   & 919.351 \ \ & \ \ 0.00120  & $95\pm10$  & ...         \\
Ly$\lambda$  & 918.129 \ \ & \ \ 0.000921 & $87\pm10$  & ...         \\
\enddata
\tablenotetext{a}{Wavelengths and oscillator strengths are from D. C. Morton 
   (2003, in preparation).}
\tablenotetext{b}{The data are best fitted by a curve of growth with 
   doppler parameter $b=13.9^{+3.1}_{-2.3}$ km s$^{-1}$ with 
   log$N$(\ion{H}{1})$=16.37^{+0.22}_{-0.14}$.}
\tablenotetext{c}{The data are best fitted by a curve of growth with 
   doppler parameter $b=8.0^{+3.5}_{-2.0}$ km s$^{-1}$ with 
   log$N$(\ion{H}{1})$=15.23^{+0.38}_{-0.22}$.}
\end{deluxetable}

\clearpage
\begin{deluxetable}{lrlcccc}
\tablecolumns{7}
\tablewidth{0pc}
\tablecaption{SUMMARY OF MEASUREMENTS: Mrk~509 \label{t4}}
\tablehead{
\colhead{} & \colhead{} & \colhead{} & 
    \multicolumn{2}{c}{Component 1}  &
    \multicolumn{2}{c}{Component 2} \\
\colhead{} & \colhead{} & \colhead{} &
      \multicolumn{2}{c}{$V_{LSR}=-240$ km~s$^{-1}$} &
      \multicolumn{2}{c}{$V_{LSR}=-300$ km s$^{-1}$} \\
\colhead{} & \colhead{$\lambda$\tablenotemark{a}} & \colhead{} &
    \colhead{$ W_{\lambda}$} & \colhead{log $N(X)$\tablenotemark{b}} &
    \colhead{$W_{\lambda}$} & \colhead{log $N(X)$\tablenotemark{b}} \\
\colhead{Species} & \colhead{(\AA)} & \colhead{$f$\tablenotemark{a}} &
    \colhead{(m\AA)} & \colhead{($N$ in cm$^{-2}$)} &
    \colhead{(m\AA)} & \colhead{($N$ in cm$^{-2}$)}  }
\startdata
\ion{H}{1}  & ...  \ \ & \ \ ...  & ... & 
       $<17.69$\tablenotemark{c} & ...  & 
       $<17.69$\tablenotemark{c}     \\
\ion{C}{2}  & 1036.337 \ \ & \ \ 0.118  & $<23$      & $<13.32$
          & $48\pm10$  & $13.68^{+0.08}_{-0.09}$      \\
          & 1334.532 \ \ & \ \ 0.128  & $<25$      & $<13.15$
          & $88\pm12$  & $13.73^{+0.05}_{-0.06}$       \\
\ion{C}{4}  & 1548.203 \ \ & \ \ 0.190  & $100\pm13$ & $13.53^{+0.05}_{-0.06}$
          & $264\pm18$ & $14.15^{+0.10}_{-0.07}$\tablenotemark{d}      \\
          & 1550.770 \ \ & \ \ 0.0948 & ...    & ...
          & $184\pm19$ & ...                          \\
\ion{N}{1}  & 1199.550 \ \ & \ \ 0.130  & $<41$      & $<13.38$
          & $<58$      & $<13.53$                      \\
\ion{N}{5}  & 1238.821 \ \ & \ \ 0.157  & $<25$      & $<13.08$
          & $<36$      & $<13.24$                                  \\
\ion{O}{1}  & 1302.169 \ \ & \ \ 0.0519 & $<22$      & $<13.47$
          & $<32$      & $<13.62$                                  \\
\ion{O}{6}  & 1031.926 \ \ & \ \ 0.133  & $77\pm6$   & $13.89^{+0.04}_{-0.04}$
          & $86\pm10$  & $13.93^{+0.04}_{-0.05}$                   \\
\ion{Si}{2} & 1193.290 \ \ & \ \ 0.585  & $<53$      & $<12.93$
          & $<72$      & $<13.09$                                  \\
\ion{Si}{3} & 1206.500 \ \ & \ \ 1.669  & $49\pm12$  & $12.44^{+0.10}_{-0.12}$
          & $194\pm19$ & $13.31^{+0.03}_{-0.04}$*                   \\
\ion{Si}{4} & 1393.755 \ \ & \ \ 0.514  & $<24$      & $<12.49$
          & $113\pm11$ & $13.53^{+0.19}_{-0.13}$\tablenotemark{d}     \\
            & 1402.770 \ \ & \ \ 0.255  & $<29$      & $<12.84$
          & $84\pm12$  & ...                         \\
\ion{S}{2}  & 1253.801 \ \ & \ \ 0.0109 & $<11$      & $<13.85$
          & $<15$      & $<14.00$                                  \\
\ion{S}{3}  & 1012.495 \ \ & \ \ 0.0442 & $<23$      & $<13.78$
          & $<33$      & $<13.93$                                  \\
\ion{Fe}{2} & 1144.938 \ \ & \ \ 0.106  & $<27$      & $<13.21$
          & $<39$      & $<13.37$
\enddata
\tablecomments{* Because some saturation is indicated in the line profile, the
    total column density may be slightly higher.}
\tablenotetext{a}{Wavelengths and oscillator strengths are from D. C. Morton
   (2003, in preparation).}
\tablenotetext{b}{Unless otherwise specified, column densities are calculated
   through the apparent optical depth method.  All upper limits are 3$\sigma$
   levels.}
\tablenotetext{c}{The upper limits on log$N$(\ion{H}{1}) are 4$\sigma$ levels
   from Sembach et al. (1995).}
\tablenotetext{d}{Column density is measured by the CoG doublet method.}

\end{deluxetable}

\begin{deluxetable}{lccccc}
\tablecolumns{6}
\tablewidth{0pc}
\tablecaption{LOGARITHMIC COLUMN DENSITY RATIOS \label{t5}}
\tablehead{
\colhead{} & \multicolumn{2}{c}{PKS 2155$-$304\tablenotemark{a}} & 
   \multicolumn{2}{c}{Mrk 509\tablenotemark{b}} & \colhead{} \\
\colhead{} & \colhead{$-140$ km s$^{-1}$} & \colhead{$-270$ km s$^{-1}$} & 
   \colhead{$-240$ km s$^{-1}$} & \colhead{$-300$ km s$^{-1}$} & \colhead{} \\
\colhead{Ratio} & \colhead{Component} & \colhead{Component} & 
   \colhead{Component} & \colhead{Component} & 
   \colhead{Complex C\tablenotemark{c}} }
\startdata
$N$(\ion{H}{1})$/N$(\ion{O}{6})   & $2.57^{+0.22}_{-0.14}$  & 
   $1.67^{+0.38}_{-0.23}$  & $<3.80$ & $<3.76$  & $6.38^{+0.05}_{-0.05}$  \\
$N$(\ion{O}{1})$/N$(\ion{O}{6})   & $<$$-0.40$  & $<$$-0.05$ & 
   $<$$-0.42$     & $<$$-0.31$    & $2.21^{+0.19}_{-0.25}$  \\ 
$N$(\ion{Si}{2})$/N$(\ion{Si}{4}) & $0.07^{+0.14}_{-0.13}$  & $<0.51$  & 
...& $<$$-0.44$   & $2.03^{+0.22}_{-0.20}$  \\
$N$(\ion{C}{2})$/N$(\ion{C}{4})   & $0.36^{+0.11}_{-0.08}$  & $<$$-0.56$  & 
   $<$$-0.38$     & $-0.44^{+0.11}_{-0.09}$ & ...        \\
$N$(\ion{C}{4})$/N$(\ion{Si}{4})  & $0.74^{+0.08}_{-0.09}$  & 
   $1.30^{+0.12}_{-0.16}$  & $>1.04$  & $0.62^{+0.21}_{-0.15}$  & 
   $0.59^{+0.12}_{-0.15}$  \\
$N$(\ion{C}{4})$/N$(\ion{N}{5})   & $>0.40$  & $>0.47$ & $>0.45$ & 
    $>0.91$     & $>0.07$                 \\
$N$(\ion{O}{6})$/N$(\ion{C}{4})   & $0.32^{+0.05}_{-0.05}$  & 
   $0.00^{+0.06}_{-0.07}$  & $0.36^{+0.06}_{-0.07}$  & $-0.22^{+0.11}_{-0.09}$     & $0.31^{+0.08}_{-0.09}$  
\enddata
\tablenotetext{a}{The mean of the column densities derived from \ion{C}{4}\ 
  $\lambda1548.2$ and $\lambda1550.8$ is used for ratios involving 
   $N$(\ion{C}{4}) for the PKS 2155$-$304 sight line.}
\tablenotetext{b}{The mean of the column densities derived from \ion{C}{2}\ 
  $\lambda1036.3$ and $\lambda1334.5$ is used for ratios involving 
   $N$(\ion{C}{2}) for the Mrk 509 sight line.}   
\tablenotetext{c}{Values for Complex C are measured along the PG~1259+593 
   sight line (Collins et al. 2003).}
\end{deluxetable}

\clearpage
\begin{deluxetable}{ccccccc}
\tablecolumns{7}
\tablewidth{0pc}
\tablecaption{MODEL CHARACTERISTICS OF HVCs TOWARD 
    PKS~2155$-$304\tablenotemark{a} \label{t6}}
\tablehead{\colhead{$Z$} & \colhead{log $N$(\ion{H}{1})} & 
   \colhead{log $U$} & \colhead{log $n_{\rm H}$} & \colhead{log $T$} & 
   \colhead{$P/k$} & \colhead{Size} \\ 
\colhead{} & \colhead{(cm$^{-2}$)} & \colhead{} & \colhead{(cm$^{-3}$)} & 
   \colhead{(K)} & \colhead{(K cm$^{-3}$)} & \colhead{(kpc)}}
\startdata
\cutinhead{$-140$ km s$^{-1}$ Component}
$-0.32$ & 16.23 & $-3.02$ & $-3.46$ & 4.02 &  8.4 & 2.6 \\
$-0.47$ & 16.37 & $-3.02$ & $-3.46$ & 4.07 &  9.4 & 4.0 \\
$-0.71$ & 16.59 & $-3.04$ & $-3.44$ & 4.14 & 11.5 & 6.9 \\
\cutinhead{$-270$ km s$^{-1}$ Component}
$-0.92$ & 15.01 & $-1.98$ & $-4.50$ & 4.34 &  1.6 &  30 \\
$-1.20$ & 15.23 & $-1.98$ & $-4.50$ & 4.38 &  1.7 &  54 \\
$-1.65$ & 15.61 & $-1.98$ & $-4.50$ & 4.42 &  1.9 & 140 \\
\enddata
\tablenotetext{a}{Results from CLOUDY photoionization model calculations.  See 
    text for a description of these models.}
\end{deluxetable}


\begin{references}
\reference{}Ballet, J., Arnaud, M., \& Rothenflug, R. 1986, \aap, 161, 12
\reference{}Bland-Hawthorn, J., Maloney, P. R. 2002, in ASP Conf. Ser. 254, 
       H$\alpha$ Distance Constraints for High Velocity Clouds in the Galactic 
       Halo, ed. J. S. Mulchaey \& J.  Stocke (San Francisco: ASP), 267
\reference{}Bland-Hawthorn, J., \& Putman, M. E. 2001, in ASP Conf. Ser. 240, 
     Gas and Galaxy Evolution, ed. J. E. Hibbard, M. Rupen, \& J. H. van Gorkom 
     (San Francisco: ASP), 369
\reference{}Blitz, L., Spergel, D. N., Teuben, P. J., Hartmann, D., \& Burton, 
      W. B. 1999, \apj, 514, 818
\reference{}Borkowski, K. J., Balbus, S. A., \& Fristrom, C. C. 1990, \apj, 355, 501
\reference{}Braun, R., \& Burton, W. B. 1999, \aap, 341, 437
\reference{}Bregman, J. N. 1980, \apj, 236, 577
\reference{}Collins, J. A., Shull, J. M., \& Giroux, M. L. 2003, \apj, 585, 336 (CSG)
\reference{}Fang, T, Marshall, H. L., Lee, J. C., Davis, D. S., \& 
            Canizares, C. R. 2002, \apj, 572, L127
\reference{}Ferland, G. J. 1996, HAZY, A Brief Introduction to CLOUDY (Univ. 
         Kentucky Dept. Astron. Internal Rep.)
\reference{}Flynn, C., \& Morell, O. 1997, \mnras, 286, 617
\reference{}Fox, A. J., Savage, B. D., Wakker, B. P., Richter, P., Sembach, K. R., \& Tripp, T. M. 2004, \apj, in press
\reference{}Giroux, M. L., \& Shull, J. M. 1997, \aj, 113, 1505
\reference{}Hartigan, P. Raymond, J., \& Hartmann, L. 1987, \apj, 316, 323
\reference{}Heckman, T. M., Norman, C. A., Strickland, D. K., \& Sembach, K. R. 2002, \apj, 577, 691
\reference{}Holweger, H. 2001, in AIP Conf. Ser. 598, Solar and Galactic Composition, ed. R. F. Wimmer-Schweingruber (New York: Springer), 23
\reference{}Hopp, U., Schulte-Ladbeck, R. E., \& Kerp, J. 2003, \mnras, 339, 33
\reference{}Indebetouw, R., \& Shull, J. M. 2004, \apj, in press
\reference{}Kraemer, S. B., et al. 2003, \apj, 582, 125
\reference{}Kravtsov, A. V., Klypin, A., \& Hoffman, Y. 2002, \apj, 571, 563
\reference{}Maloney, P. R., \& Putman, M. E. 2003, \apj, 589, 270
\reference{}Mathur, S., Weinberg, D. H., \& Chen, X. 2003, \apj, 582, 82
\reference{}Moos, H. W. 2000, \apj, 538, L1  
\reference{}Nicastro, F., et al. 2002, \apj, 573, 157
\reference{}Nicastro, F., et al. 2003, Nature, 421, 719 
\reference{}Putman, M. E., Bland-Hawthorn, J., Veilleux, S., Gibson, B. K., 
  Freeman, K. C., \& Maloney, P. R. 2003, \apj, 597, 948
\reference{}Rasmussen, A., Kahn, S. M., \& Paerels, F. 2003, in The IGM/Galaxy
  Connection, (Dordrecht: Kluwer Publ.), ed. J. Rosenberg \& M. Putman, 109
\reference{}Reynolds, R. J. 1993, in AIP Conf. Proc. 278, Back to the Galaxy, 
     ed. S. S. Holt \& F. Verter (New York: AIP), 156
\reference{}Richter, P., et al. 2001, \apj, 559, 318
\reference{}Savage, B. D., \& Sembach, K. R. 1991, \apj, 379, 245
\reference{}Sembach, K. R., et al. 2003, \apjs, 146, 165 
\reference{}Sembach, K. R., \& Savage, B. D. 1992, \apjs, 83, 147 
\reference{}Sembach, K. R., Savage, B. D., Lu, L., \& Murphy, E. M. 1995, \apj,                 451, 616
\reference{}Sembach, K. R., Savage, B. D., \& Tripp, T. M. 1997, \apj, 480, 216
\reference{}Sembach, K. R., Savage, B. D., Lu, L., \& Murphy, E. M. 1999, \apj,
    515, 108 (S99)
\reference{}Shapiro, P. R., \& Field, G. B. 1976, \apj, 205, 762
\reference{}Shull, J. M., Tumlinson, J., \& Giroux, M. L. 2003, \apj, 594, L107
\reference{}Simon, J. D., \& Blitz, L. 2002, \apj, 574, 726
\reference{}Slavin, J. D., \& Cox, D. P. 1993, \apj, 417, 187
\reference{}Slavin, J. D., Shull, J. M., \& Begelman, M. C. 1993, \apj, 407, 83
\reference{}Songaila, A., \& Cowie, L. L. 1996, \aj, 112, 335 (SC)  
\reference{}Sutherland, R. S., \& Dopita, M. A. 1993, \apjs, 88, 253
\reference{}Telfer, R. C., Zheng, W., Kriss, G. A., \& Davidsen, A. F. 2002, 
     \apj, 565, 773
\reference{}Tripp, T. M., et al. 2002, \apj, 575, 697
\reference{}Tripp, T. M., et al. 2003, \aj, 125, 3122 
\reference{}Tufte, S. L., Wilson, J. D., Madsen, G. J., Haffner, L. M., 
    Reynolds, R. J 2002, \apj, 572, L153
\reference{}Wakker, B. P., et al. 1999, \nat, 402, 388
\reference{}Wakker, B. P., Kalberla, P. M. W., van Woerden, H., de Boer, K. S., 
     \& Putman, M. E. 2001, \apjs, 136, 537 
\reference{}Wakker, B. P., \& van Woerden, H. 1997, \araa, 35, 217
\reference{}Weiner, B. J., Vogel, S. N., \& Williams, T. B. 2002, in ASP Conf. Ser.
      254, Extragalactic Gas at Low Redshift, ed. J. S. Mulchaey \& J. Stocke 
      (San Francisco: ASP), 256
\reference{}Zheng, W., Kriss, G. A., Telfer, R. C., Grimes, J. P., \& Davidsen, 
     A. F. 1997, \apj, 475, 469
\reference{}Zwaan, M. A. 2001, \mnras, 325, 1142
\end{references}
\end{document}